\documentclass[aps,prl,showpacs,twocolumn]{revtex4}
\usepackage{graphicx}
\begin{document}
\newcommand {\be}{\begin{equation}}
\newcommand {\ee}{\end{equation}}
\newcommand {\bea}{\begin{eqnarray}}
\newcommand {\eea}{\end{eqnarray}}
\newcommand {\nn}{\nonumber}

\title{ Quasiperiodic Heisenberg antiferromagnets in two dimensions}

\author{ Anuradha Jagannathan}
\affiliation{Laboratoire de Physique des Solides, CNRS-UMR 8502, Universit\'e
Paris-Sud, 91405 Orsay, France }

\date{\today}

\begin{abstract}
We describe some of the properties of 2d quantum quasiperiodic antiferromagnets as reported in studies that have been carried out in the last decade. Many results have been obtained for perfectly ordered as well as for disordered two dimensional bipartite quasiperiodic tilings. The theoretical methods used include spin wave theory, and renormalization group along with Quantum Monte Carlo simulations. These methods all show that the ground state of these unfrustrated antiferromagnets have N\'eel type order but with a highly complex spatial distribution of local staggered magnetization. The ground state properties, excitation energies and spatial dependence, structure factor, and local susceptibilities are presented and discussed. The effects of introducing geometrical disorder on the magnetic properties are discussed.

\end{abstract}
\maketitle
\setcounter{secnumdepth}{1}
\setcounter{secnumdepth}{2}
\section{Introduction}
This article reviews properties of the Heisenberg model for a quantum spins in a quasiperiodic antiferromagnet. The quasiperiodic structures considered are two well-known models of quasicrystals, the Penrose \cite{pen,gard} and the Ammann \cite{ammann} tilings. Quasicrystals are highly organized states of matter (see \cite{stein}), with long range positional order that is a generalization of the usual translation invariance found in crystals. In a crystal, one can identify a basic motif which is replicated at regular distances, whereas in a quasicrystal, one can discern patterns which repeat in space, in aperiodic fashion. The simplest example of an aperiodic structure is a one-dimensional array of atoms with two characteristic spacings, say L and S (for long and short respectively) in which the spacings follow the Fibonacci sequence, ...LSLSLLSLS...according to a perfectly deterministic algorithm. Due to the complexity of aperiodic structures compared with crystals, they have sometimes been considered as being in some sense closer to disordered structures insofar as their physical properties are concerned. As we will see, on the contrary, the symmetries present in quasicrystals can lead to novel and striking features that are not found in a crystal, much less a disordered material.

The presence of Bragg peaks in the structure factor is one defining characteristic of a perfectly ordered quasiperiodic structure. Unlike crystals, where Bragg peaks occur at regular intervals determined by a set of $d$ reciprocal lattice vectors (where $d$ is the dimensionality), the Bragg peaks of a $d$-dimensional quasicrystal are described with the help of $n>d$ reciprocal lattice vectors. The peaks are not regularly spaced, and their intensities cannot be simply expressed as a function of the position. In principle, the reciprocal space is densely filled with peaks most of them of vanishingly small intensity, so that in an actual experiment, one sees a large but not infinite number of peaks that have an amplitude greater than some experimentally observable threshold value. For real quasicrystal samples, moreover, the experimentally observed diffraction spots have finite widths due to instrumental resolution and sample imperfections. The widths of the spots indicate the correlation length, or distance out to which quasiperiodic order is present, and this is comparable to those observed in crystalline materials, of the order of microns in the best samples available. Since the constraints imposed by translational invariance are not present for the quasicrystal, the usual crystallographic rules forbidding certain rotational symmetries such as 5-fold symmetry, do not apply. The observation of five-fold symmetry of the diffraction pattern led to the experimental discovery by Schechtman et al \cite{schecht} of the first quasicrystalline alloy, composed of aluminum, copper and silicon, in 1984. Other crystallographically forbidden symmetries, such as 8-fold, 10- fold or twelve-fold symmetries have also been discovered and known quasicrystal alloys now number in the hundreds.

In the above we have assumed that the structure is perfectly ordered, however, many quasicrystals are thought to be intrinsically disordered, and better described by random tiling models rather than the above-mentioned deterministic structures. Random tilings \cite{elserprl} are composed of the same units as the deterministic tilings, but do not have the hierarchical invariance of the perfect systems. A perfect structure can be disordered, for example, by allowing for random local phason flips -- when an atom has the freedom to flip between two alternative positions without affecting the local bond lengths and angles. Here we are interested in the effect of this type of disorder on the antiferromagnetic ground state properties.

\subsection{Magnetic quasicrystals and the Heisenberg model}
The Heisenberg model can be used to discuss quasiperiodic alloys such as the ZnMgR alloys (where R stands for a rare earth Ho,Ti,Dy,..), where the magnetism is due to localized f-orbitals. Properties of transition metal alloy systems such as AlMnPd which have itinerant magnetic moments will not be discussed here (see review by Hippert et al in \cite{belin}). Magnetization measurements for the icosahedral ZnMgR and CdMgR compounds show that the predominant interactions are antiferromagnetic \cite{hattori,char1,fisher,sato1}. Neutron scattering experiments \cite{char2,islam,sato,sato2} show that although short magnetic correlations set in at low temperature, there is no long range antiferromagnetic order down to the lowest temperature studied. Instead, in some of these compounds, a spin-glass type low temperature phase and out-of-equilibrium phenomena was found. These phenomena arise due to magnetic frustration as well as, no doubt, to the presence of some amount of disorder. Crystal field effects have been argued to be important as well \cite{fisher}. The experimental data raise a number of interesting theoretical questions. What is the ground state of a quasiperiodic antiferromagnet  and how is it different from that of periodic antiferromagnets ? What is the equivalent of spin wave modes or magnons in a quasicrystal ? What are the effects of disorder ? Are there novel frustration-induced phenomena ?

These questions are difficult to answer, especially as regards the effects of disorder and frustration, whose effects remain insufficiently understood even in periodic structures, despite the many studies devoted to them. We will here only consider the simplest situation: that of systems without frustration -- those in which the spins have a ground state with a N\'eel type (up-down) ordering. The most tractable of quasiperiodic models is the Fibonacci chain, and the behavior of quantum spins on
quasiperiodic chains been considered by several authors.
Quantum
spin chains have been analyzed using renormalization schemes
\cite{herm,hida,hida2} based on the inflation symmetry of these chains.
Using a mapping to fermionic models and techniques of bosonization, results have been obtained
 for global properties such as the magnetization as a
function of external field, and the spectral gaps for a variety of
different quasiperiodic sequences \cite{vid}. In contrast to two and higher dimensions, however, there is no symmetry breaking in one dimension.

Two and three dimensions are interesting because long range magnetic ordering can occur at zero temperature (destroyed at finite temperature in the case of $d$=2). In two dimensions, the low connectivity leads to larger quantum effects, compared to three dimensions. In addition, some analytical calculations are possible for two dimensions, and numerical calculations can be more easily carried for large systems compared to three dimensional structures. We will be interested in the quantum fluctuations in the quasicrystal, which are maximal when the spin S of the atoms is small. We will first describe the antiferromagnetic ground state of perfect quasicrystals in some detail, before going on to the case of disordered quasicrystals. It should be noted that this quantum limit is not expected to be pertinent in the rare earth quasicrystals where the spin S is on the contrary quite large, and the classical Heisenberg model should suffice. Competing nearest-neighbor and next-nearest neighbor interactions in these systems would lead in these systems to complex behavior at low temperatures. Such models are outside the scope of the current review, but have been studied for classical spin systems in two dimensions \cite{godluc,duneau,ledue,vedmed,vedmed2} and three dimensions \cite{matsu,sato2}.

In recent years, new progress has been made in the attempt to fabricate quasicrystalline structures by deposition of atoms on clean surfaces of perfect quasicrystals. This is an exciting development, as two dimensional structures based on one or two adatom species are easier to model theoretically, making it possible to study in a controlled way the consequences of long range quasiperiodic geometric ordering on the electronic and magnetic properties. There has been much recent work on growing thin films on quasicrystalline surfaces that display quasiperiodic order out to distances of hundreds of nanometers. Characterization of these structures, and studies of the interplay of geometrical and electronic properties of these films are ongoing ( as reviewed in \cite{mcgrath}). Magnetic properties of atoms in surface layers have not yet been investigated, and the realization of a Heisenberg model in thin quasiperiodic films represents an experimental challenge.

Sec.II presents an overview of the two tilings with some definitions and geometrical properties. Sec.III provides an introduction to properties of the Heisenberg model for quantum spins. Sec.IV introduces the models that are studied, whose results are then discussed in Secs.V and VI. Sec. VII discusses disordered models. Sec.VIII presents the conclusions.

\section{Tiling models}
 Investigations of quasicrystal properties often begin with studies based on quasiperiodic tilings. In the way that the Bravais lattices are used as templates for crystals, quasiperiodic tilings are useful as a starting point for a description of real alloys. Depending on the dimensionality and symmetries of the tilings, different tilings help to classify the various members of this large class of materials. In three dimensions, the icosahedral tiling (also called 3D Penrose tiling) \cite{levine,elser,kalu,dunkatz} is the basis for most models of realistic atomic structures of quasicrystals. In one dimension, the simplest example of a tiling is the Fibonacci ``tiling" or chain already referred to above. The best known tiling is, probably, the 2D Penrose tiling \cite{pen,gard}, which was introduced before the discovery of quasicrystals. It is built out of elementary rhombus-shaped tiles that come in two shapes - the thick and the thin - each of which can occur in five orientations as can be seen in Fig. 1, leading to the five-fold symmetry of the diffraction pattern. This structure gives rise to delta-function peaks in its structure factor (the modulus squared of the Fourier transform of a crystal formed by, say, placing a point mass on every vertex). The octagonal or Ammann-Beenker tiling is another often studied two-dimensional tiling. It is composed of two kinds of tiles: the square and the 45$^\circ$ rhombus -- the tiles shown in Fig.\ref{octafig.fig}.

\begin{figure}[h] \begin{center}
\includegraphics[width=8cm]{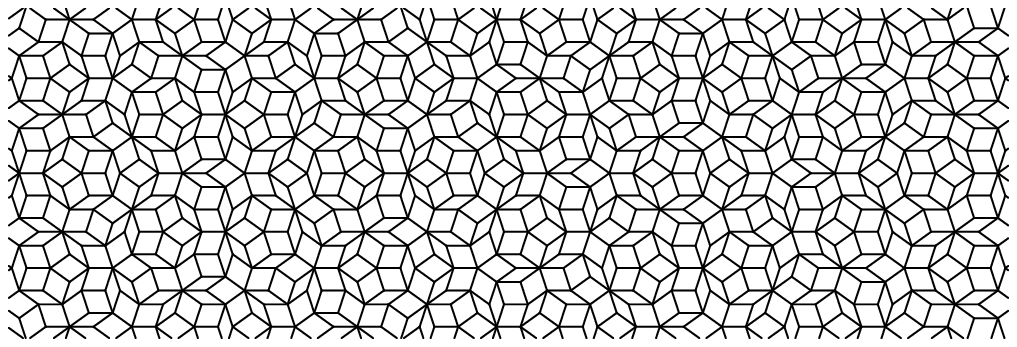}
\caption{\label{fig:tiling} A finite patch of the perfect Penrose tiling.}
\end{center}\end{figure}

\subsection{Symmetries} One readily observable
feature of a tiling such as the one in Fig.\ref{fig:tiling} is that, despite the absence of perfect translational
invariance, patterns of any given size frequently repeat (this ``local isomorphism" property is explained by Baake in \cite{janmoss}). The mean repetition distance of a pattern of
linear size R is of the order of R, a property that replaces the strict
translational invariance of crystalline structures.

Next, we discuss the rotational symmetry present in these tilings. This holds in the same weaker sense than in crystals, as we saw with the repetitivity property. The rotational invariance is true in the following sense -- any given configuration of tiles of arbitrary size R will be found to occur with equal probability in each of its $n$ rotated forms, where $n=5$ for the Penrose rhombus tiling and $n=8$ for the octagonal tiling.

\begin{figure}[h]
\begin{center}
\includegraphics[scale=0.4]{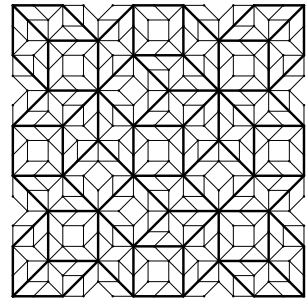}
\caption{ Inflation of the octagonal tiling showing the original tiling (thin lines) and the inflated one (thick lines) }
\label{octafig.fig}
\end{center}
\end{figure}

The tilings possess a
hierarchical symmetry under discrete scale transformations. These are the so-called inflation and
deflation transformations illustrated in Fig.\ref{octafig.fig}. Inflation (deflation) is a reversible operation of change of scale which
consists of combining (subdividing) each of the tiles into larger (smaller) tiles of the same shape. In the case of the Penrose tiling, the tiles of the new system thus obtained are
 bigger (smaller) by a factor
$\tau=\frac{\sqrt{5}+1}{2}$ (the golden mean). In the case of the octagonal tiling, the edge length is increased by the number
$\lambda=\sqrt{2}+1$ (which goes by the name of silver mean). No new environments are created
or destroyed in the process of inflation or deflation. The original and the inflated tilings are equivalent in the sense that for any given configuration of tiles in one of the tilings, one can always find
an exactly identical region (up to a length change) in the other tiling.

The self-similarity of quasiperiodic tilings under inflation-deflation operations is an important symmetry property, and can be used in defining renormalization transformations. In the one-dimensional case, this property has been exploited to calculate properties of spin models on the Fibonacci chain and other so-called metallic mean chains \cite{hida}. There are fewer results available in the case of two dimensional tilings. A renormalization argument was used in \cite{godluc} to map out a phase diagram for a frustrated Ising model on the Penrose tiling.  In the case of the octagonal tiling, an approximate renormalization group calculation leads to a nontrivial fixed point solution for the antiferromagnetic ground state, as will be discussed in the next section \cite{jag,jagrg}.

The last symmetry property that we will mention has to do with the bipartite character of the tilings. Quadrilateral (or more precisely, rhombus) based
structures such as the ones shown in Fig.\ref{fig:tiling} are constituted of two sublattices. Less obviously, it can be shown that for the infinite quasiperiodic tiling, the two sublattices are exactly
equivalent, as regards the number and the nature of the vertices (see below).

\subsection{Local environments}
\begin{figure}[h] \begin{center}
\includegraphics[width=6cm]{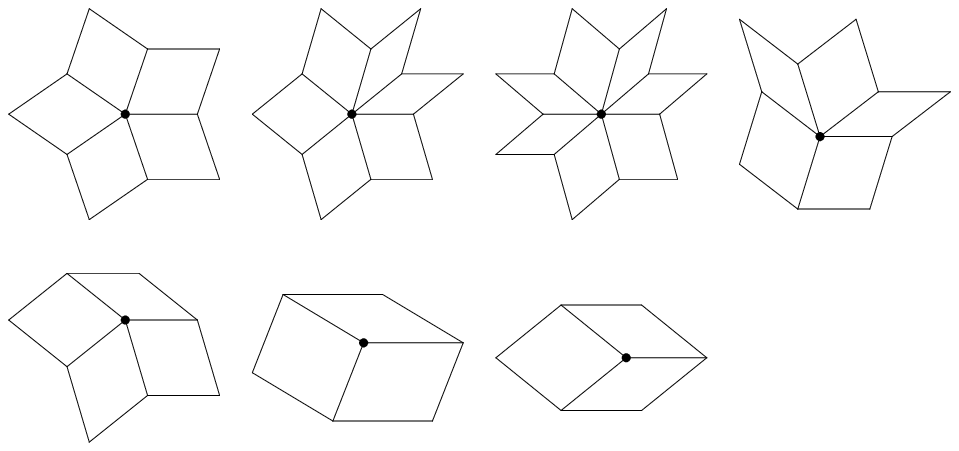}\hskip 1cm
\includegraphics[scale=0.40]{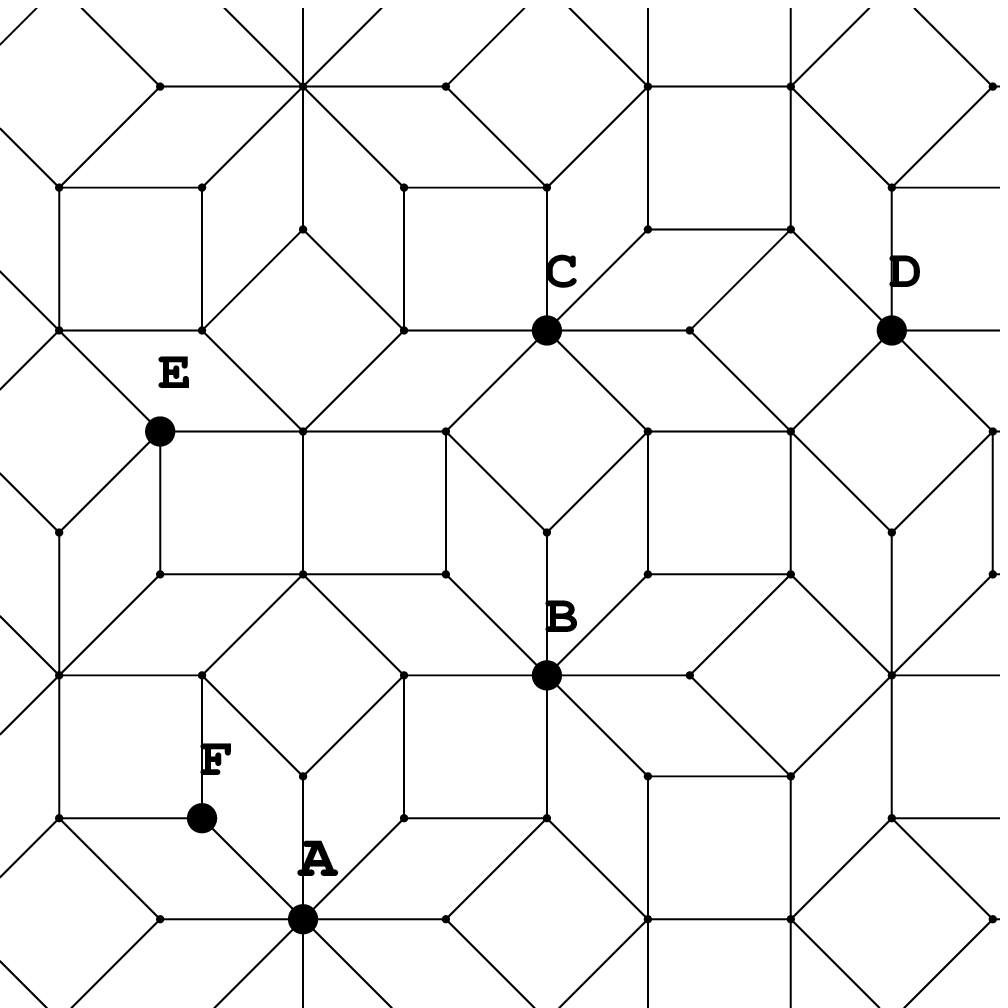}
\caption{\label{fig:starshapes} (top) Local environments in the Penrose tiling (Reproduced from \cite{sza3}). (bottom)A portion of the octagonal tiling showing
the six different nearest neighbor environments A,B,...,F (Reproduced from \cite{jagrg})}
\end{center}\end{figure}

The upper figure in Fig.\ref{fig:starshapes} shows the different kinds of vertices or local environments present in
the Penrose rhombus tiling. The seven local environments correspond to site
coordination numbers (number of nearest neighbors) ranging from 3 to 7. The frequencies of each of the vertices can be calculated in terms of $\tau$. The average value $\overline{z}$ being exactly 4, as it is in the octagonal tiling.
The lower figure shows a portion of the octagonal tiling labelling six different types of vertices, with the coordination numbers ranging from 3 to 8. Again, the frequencies of each kind of vertex can be exactly found, in terms of the number $\lambda$. The transformation of vertices under inflation or deflation on the Penrose and the octagonal tiling will not be described here in detail. We will return to this point in the section on renormalization group for the octagonal tiling.

\subsection{Tilings as projections from higher dimensions}
 There are many ways to obtain quasiperiodic tilings. It is outside the scope of this review to describe in detail the methods that exist.
 In this section we will briefly outline the cut-and-project method used to obtain samples of the Penrose tiling in our calculations, since it will turn out to be helpful for understanding the results that will be obtained for the Heisenberg spin model.

\begin{figure}[ht]
\begin{center}
\includegraphics[scale=0.60]{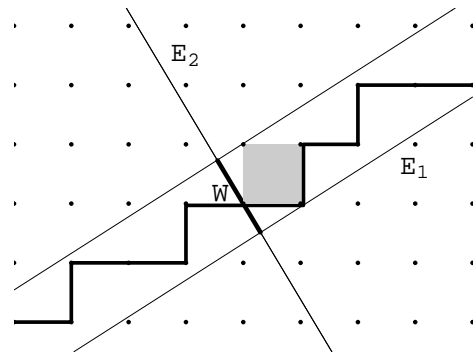}
\vspace{.2cm} \caption{Obtaining a piece of the Fibonacci chain by the cut and project method. The bold line indicates the edges which are selected and form the Fibonacci chain after projection onto the physical line $E_1$. The perpendicular space axis is denoted $E_2$. } \label{projfig.fig}
\end{center}
\end{figure}

 The Penrose tiling (PT) and octagonal tiling (OT) can be obtained by projecting
a subset of vertices of a D dimensional cubic lattice \cite{dunaudier}, where $D=5$ for the Penrose tiling and $D=4$ in the case of the octagonal tiling.  In the projection method, the $d$-dimensional tiling is obtained by projecting a subset of points of a D-dimensional periodic lattice (where $D>d$). The value of D and the selection of the subset of points to be projected are found by considering the symmetries one would like the tiling to possess. The Fibonacci chain, for example, can be obtained by projecting all the points lying within an infinite strip of a square lattice of slope $1/\tau$, as shown in Fig.\ref{projfig.fig}. Inside the strip, whose width corresponds to one unit square, neighboring vertices are connected by edges whose projections give rise to the L and S line segments that make up the tiling. The figure shows by a thick line the set of edges that are selected for projection. The irrational slope of the strip ensures that the projected pattern never repeats.

If, now, the irrational number were to be replaced by a nearby rational, for example one of the approximants of the golden mean, $1/\tau_k = F_k/F_{k+1}$, the strip becomes commensurate with the square lattice. It is easy to check, using the recursion relation for the Fibonacci numbers, that the projected pattern would in that case repeat after a distance $L_k = F_{k+2}$. This type of so-called $approximant$ of the Fibonacci chain is useful for numerical calculations, where one considers finite systems with periodic boundary conditions, and attempts to extrapolate to the quasiperiodic limit where $k\rightarrow \infty$. The infinite chain can be transformed into itself under inflation (or deflation) -- in an inflation, tiles are grouped into new bigger tiles $L'$ and $S'$ ($LS \rightarrow L'$ and $L\rightarrow S'$). The new chain obtained in this way has the same properties as the old, with the difference being a scale factor $\tau$.

In an analogous way, one can obtain the octagonal tiling by projecting lattice points lying inside an infinite cylinder in four dimensional space. The orientation of the cylinder is given in terms of the so-called ``silver mean" $\lambda=(\sqrt{2}+1)$. It is this irrational number that determines the shapes of the tiles after projection, and it determines the scale of the inflation/deflation transformation of the octagonal tiling. The Penrose tiling in two or three dimensions, similarly, may be obtained by projection, and the irrational number $\tau$ appears again in the matrices used for projection and in the tile-rescaling transformations. As in the case of the Fibonacci chain, it is possible to obtain approximants of the octagonal tiling. These are finite square tilings that can be periodically continued in the plane, and whose period can be increased progressively by considering successive approximants of $\lambda$ \cite{ogu}. Similarly, for numerical studies of the Penrose tiling antiferromagnet, one can work with the so-called Taylor approximants obtained by a generalization of the method described by Duneau and Audier in \cite{dunaudier}.

When the vertices of the tiling are projected not onto the physical plane, but instead into perpendicular space, they fall into a compact region (the ``selection window"). In the case of the Fibonacci chain, one sees that points project onto the line segment labelled by W in Fig.\ref{projfig.fig}. In the case of the octagonal tiling, all the projected points fall into an octagonal shaped domain in the two dimensional perpendicular space.  The vertices of the Penrose tiling can be similarly projected into a three dimensional perpendicular space, where they lie in pentagon-shaped plane domains as seen in Fig.\ref{perpsp.fig}. In each of these cases, sites that have the same local environment are found grouped together in compact domains. In other words, the perpendicular space representation allows us to separate sites not only according to their nearest neighbor configurations (coordination number) but also out to further distances.
We will later exploit this useful property of the
tilings in order to represent the complex spatial structure of the antiferromagnetic
ground state in a compact way.

The codimension, or the difference $D-d$, is a rough measure of the degree of complexity of the quasiperiodic structure obtained in the projection method. The octagonal tiling has a co-dimension of 2. The Penrose tiling can be obtained from a projection down from $D=5$ (one can bring this down to $D=4$, but with a less transparent procedure), and has a slightly larger set of local environments than the octagonal tiling. If one projects a three dimensional cubic lattice onto a suitable plane one obtains the two-dimensional generalized Rauzy tiling \cite{vidmoss}, which is of co-dimension 1. This last example is close to being an integrable system, as shown by the analyses of its spectral properties for the tight-binding electron problem \cite{review}. Going to higher dimensions and codimensions $>2$, one has the entire family of rhombus tilings described in \cite{vidal2}. These have increasingly complex environments as the codimension gets larger.

\section{Introduction to quantum antiferromagnets}
The Heisenberg model expresses the energy as a function of the orientation of the local moments of the atoms, and can be derived as an effective low energy model for interacting electron systems. The model considers a set of spins $\bf{S}_i$, situated on the sites $i$ of a lattice or tiling and interacting via the Hamiltonian $H=\sum J_{ij} S_i.S_j$. The signs and the magnitudes of the spin-spin couplings $J_{ij}$ depend on the  physical and chemical details of the structure, and these variations give rise to the large variety of magnetic phases observed in solids.

\begin{figure}[h]
\begin{center}
\includegraphics[scale=0.25]{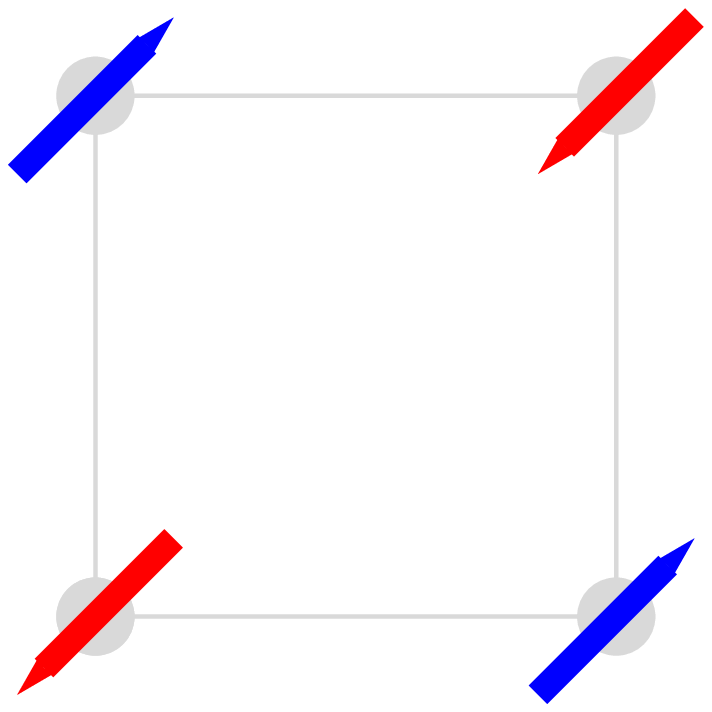}
\includegraphics[scale=0.5]{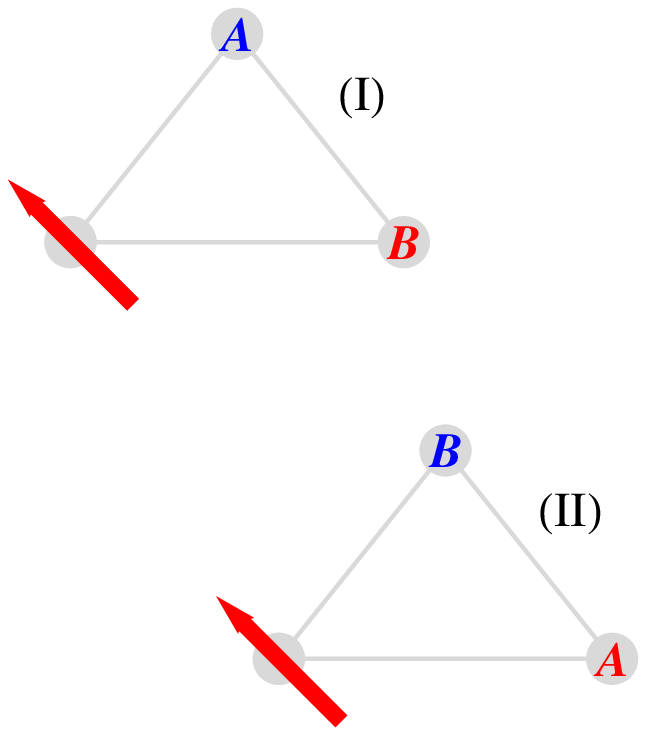}
\caption{ Examples of minimal energy spin configurations: in an unfrustrated (left) square where the spins alternate in direction,  and a frustrated (right) triangle where the energy is minimized by two different configurations I and II (A and B are two vectors oriented at $\pm 120^\circ$ angles with respect to the third spin).}
\label{frustr.fig}
\end{center}
\end{figure}

The simplest case, ferromagnetism, corresponds to one in which all of the couplings have negative sign $J_{ij}<0$ -- it is then energetically favorable to have spins aligned parallel to each other. This results in a unique ground state configuration (upto global rotations). The resulting ground state has a simple description in classical as well as in quantum theory. This is not true in the antiferromagnetic case, when the coupling $J$ has a positive sign, as we now explain. To begin with, a new phenomenon called frustration can now occur. To illustrate this, let us consider antiferromagnets in which the coupling between nearest neighbor spins is $J>0$. For simplicity, let us assume other couplings can be neglected. Unfrustrated systems are ones which comprise a sublattice A in which all spins point "up" (with respect to the axis of symmetry breaking) and a sublattice B in which spins point "down", in such a way that all neighbors of A-sublattice sites belong in the B-sublattice, and vice versa.  Fig.\ref{frustr.fig} shows one example of an unfrustrated and an frustrated loop. The unfrustrated loop has one possible minimal energy configuration (upto global transformations). The frustrated triangular loop minimizes its energy when the three spins are aligned at 120$^\circ$ angles with respect to each other, and this is possible in two different ways. In a large system formed from many such units, as a result, frustration can sometimes lead to very highly degenerate ground states, and such systems do not show a transition down to very low temperatures ($T<<J$).

Unfrustrated antiferromagnets are expected to have a unique ground state. For classical systems the ground state is characterized by broken rotational symmetry, with the spins ordering in an up-down ``Neel'' order.  The order parameter, called the "staggered magnetization", is defined by $m_s(T)=\vert S_{iz}\vert $ where $z$ is the axis of symmetry breaking. In three dimensions, $m_s$ is zero above a critical temperature $T_c$, while below $T_c$, it grows as the temperature is decreased towards the maximum value possible. In one and two dimensions, $m_s(T)=0$ at any nonzero temperature (Mermin-Wagner-Hohenberg theorem).

At $T=0$ and for classical spins in any dimension, the order parameter is $m_s^{cl}(T=0)=S$,  where $S$ is the spin length. In contrast, in the case of quantum spins, the order parameter $m_s^{cl}(T=0)$ is reduced below this value, due to quantum fluctuations, which exist even at $T=0$. Quantum fluctuations become more important as the spin length decreases, and are most important for $S=\frac{1}{2}$. The magnitude of the order parameter depends on the type of lattice, and it can vanish if the connectivity is sufficiently low, as happens in the one dimensional chain \cite{bethe}. This is of course also true for the one-dimensional quasiperiodic systems: such as the Fibonacci chain Heisenberg antiferromagnet \cite{herm,hida}. Two dimensional models were an open question for a while. Interest in the two dimensional Heisenberg antiferromagnet rose to a peak in the mid-eighties with the discovery of the high-Tc superconductors. Many different analytical and numerical approaches \cite{manou} have been brought to bear on the square lattice and other periodic antiferromagnets. The challenges faced are due to the fact that unlike ferromagnets, the ground state of the antiferromagnet is impossible to write explicitly for even a simple case such as the square lattice. There is close competition for ground state status between the antiferromagnetic state and a state with no magnetic long range order such as Anderson's RVB -- for resonating valence bond -- state\cite{anderson1} which resurfaced in the context of high-$T_c$ superconductor physics \cite{anderson2}.

The results of these studies have settled the controversy for most of the community, and it is accepted that there is indeed long range antiferromagnetic order for spins $S=\frac{1}{2}$ at $T=0$ in the two important unfrustrated two dimensional periodic lattices, namely the square and honeycomb.   As coordination number decreases, quantum fluctuations increase. This is illustrated by the fact that the order parameter is smaller in the honeycomb lattice ($m_s=0.235$\cite{rieg}) compared to the square lattice ($m_s=0.307$\cite{sand}).

Our aim here is to study the interplay of magnetism with the lattice geometry in an unfrustrated  quasiperiodic antiferromagnets in two dimensions by considering the Heisenberg model on the octagonal and Penrose tilings. We will consider both perfect and phason-disordered (see last section) samples of these tilings. Other tilings in two dimensions that it would be interesting to study are the 2d Rauzy tiling \cite{vidmoss}, which is closer to the square lattice than the previous two, and, at the opposite extreme in terms of complexity of structure, the high codimension random rhombus tilings introduced by Destainville et al \cite{vidal2}. The labyrinth \cite{siremoss} tiling, which has the connectivity of the square lattice, and shares the same ground state properties if the couplings are assumed to be uniform in value.

\section{Models of quasiperiodic antiferromagnets and methods of solution.}
We consider that spins are coupled only to their nearest neighbors, as defined by the edges of the tiling. We will assume all couplings to have the same value J. The spatial complexity of the solutions of this model and nontrivial magnetic ground state and spin excitations arise, thus, solely due to the complexity of  the quasicrystalline structure in real space. The Hamiltonian takes the form

\bea
H=\sum_{\langle i,j\rangle} J_{ij} \vec{S}_i.\vec{S}_j
\eea
where $\vec{S}_i$ is a spin variable on the site $i$, situated at position $\vec{r}_i$. The antiferromagnetic couplings $J_{ij}=J>0$ for all pairs of sites $i$ and $j$ that are linked by an edge, and zero otherwise. We will consider the case where the quantum number $S=\frac{1}{2}$, the case of greatest theoretical interest, as
quantum effects are expected to be strongest. The systems we consider are bipartite, i.e. each site belongs to one of two
sublattices A and B, and the $J$-terms couple a spin on sublattice A
to spins lying on sublattice B and vice-versa. This property ensures that there is no
frustration, i.e., if one considers classical spin variables, the
ground state is one for which all bonds are ``satisfied" -- with all
the A-sublattice spins pointing in one direction and B-sublattice
spins pointing in the opposite direction.

This quasiperiodic unfrustrated Heisenberg antiferromagnet is expected to have
long range order at zero temperature, like the unfrustrated periodic two dimensional systems already mentioned.
In the ground state, rotational symmetry is broken, and
spins acquire a non-zero magnetization in the $z$-direction. In the quasicrystal, one can define a local staggered magnetization for every site

\bea
m_{si} = \langle \epsilon_i S_{iz}\rangle
\eea
where $\epsilon_i =\pm 1$ depending on whether
the site $i$ is on the A or B sublattice. In a translationally invariant system, the local staggered magnetization is independent of $i$, while in the quasicrystal each site is different from all the others and we expect, in general, a distribution of values of $m_{si}$. Due to the symmetries of the quasicrystal, we will see that this distribution possesses very unique spatial properties. The methods that have been used to analyse the ground state properties are described below.

\subsection{The Heisenberg star cluster}
\begin{figure}[h]
\begin{center}
\includegraphics[width=3.0cm,angle=0]{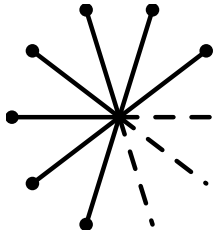}
\vspace{.1cm} \caption{Heisenberg star with $z$ neighboring spins. }
\label{hstar.fig}
\end{center}
\end{figure}

We begin with a description of the energy of a simple cluster of spins, the Heisenberg stars. A central spin $\bf{S}_0$ is coupled to $z$ neighbors $\bf{S}_j$ as shown in Fig.\ref{hstar.fig}. The corresponding Hamiltonian, $H_{HS}(z) = J\sum_{j=1}^{z} {\bf S}_0 \cdot {\bf S}_j$, can be exactly solved. The result for the lowest energy level of the cluster is

\begin{equation}
\epsilon^{(0)}(z) = -JS (zS+1) = -J \frac{(z+2)}{4}
\label{heis.eq}
\end{equation}
where the second equality holds for the case of $S=\frac{1}{2}$. The superscript 0 denotes that this is the energy of an isolated cluster. This expression of the cluster energy is the first step, as we will explain later, for a recursive calculation of the energy in an extended structure. The degeneracy of the ground state manifold is
$2S_{tot}+1 = z$, rotational symmetry is not broken, and the value of the staggered magnetization is zero for all the spins. One can use the average value of the spin-spin correlation function to define a quantity that has the same dimensions as the magnetization squared, namely
\bea
\mu^2(z) &=& \frac{1}{z+1}(\langle{\bf S}_0.{\bf S}_0 + \sum_{j=1}^{z} \vert{\bf S}_0 \cdot {\bf S}_j\rangle_0\vert) \nonumber \\
&=& \frac{1}{z+1}(S(S+1)- \epsilon^{(0)}(z)/J) \nonumber \\
&=&  \sqrt{\frac{5 + z}{4 (z+1)}}
\label{heismag.eq}
\eea
where  $\langle..\rangle_0$ are the expectation values evaluated in the ensemble of ground states, and where the last equality holds for the case of $S=\frac{1}{2}$. Note that for $z$ large, $\mu$ tends to the classical value of the magnetization, $\mu (z\rightarrow\infty)\rightarrow S$ and also that the quantum corrections to $\mu$  decrease with increasing $z$.

\subsection{Quantum Monte Carlo calculations}
Relatively large -- of the order of several thousand spins -- samples can be studied using
Stochastic Series Expansion Quantum Monte Carlo algorithm\cite{sand}
(SSE-QMC). The first calculations were carried out \cite{wess} for the octagonal tiling antiferromagnet. The ground state energy per site (GSE) of the quasicrystal was determined by repeating calculations for a series of samples of finite size N (number of sites). The finite size scaling argument used for periodic systems predicts that the corrections to the ground state energy decrease as $N^{-3/2}$ for two-dimensional lattices \cite{fss}. The argument used can be generalized to the quasiperiodic case as well, and indeed the numerical results show the expected scaling. This allows one to extract the infinite size limit of the ground state energy. The value of the GSE found for the octagonal tiling \cite{wess} is $E_{gs,OT}^{QMC} =-0.6581(1)$, to be compared with the square lattice value \cite{sand} of $E_{gs,sq}^{QMC}=-0.66943$. For the Penrose tiling, the GSE found by QMC \cite{sza} is $E_{gs,PT}^{QMC}=-0.6529(1)$. All energies are expressed in units of $J$, the exchange coupling. One sees that the ground state energy of the square lattice lies below that of the octagonal tiling which lies below that of the Penrose tiling. As the classical values for all three systems is exactly equal, $E_{gs}^{cl}=-0.5$, the differences are due to the difference in quantum fluctuations in these three types of antiferromagnets. This point will be discussed again in the section devoted to the energy spectrum in linear spin wave theory.

\begin{figure}[h]
\includegraphics[scale=0.4]{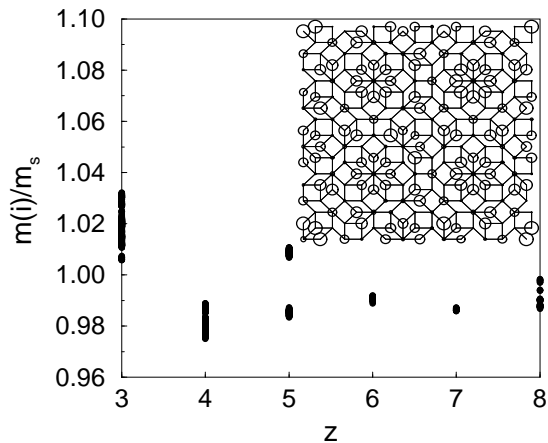}
\vspace{0.0cm}
\caption{
The normalized staggered moments
 plotted against coordination number $z$ (calculation for N=1393 approximant of the OT).
The inset shows the spatial variation of the staggered magnetization
where the radii of the circles correspond to the size of the local
staggered magnetizations.(Reproduced from \cite{wess})
}
\label{staggmagQMC.fig}
\end{figure}

The staggered magnetizations cannot be directly measured, as one is dealing with finite systems, for which there is no symmetry breaking. The expectation value $\langle S_{iz}\rangle $ giving the staggered moment on site $i$ is zero. In periodic lattices, it is nevertheless possible to obtain {\it estimates} of the values of the local staggered moment at each site by summing the pair correlation function over all pairs of spins. This estimator becomes exact in the thermodynamic limit, $N \rightarrow
\infty$. Generalizing to the tiling, one defines
\bea
\mu^{QMC}(i)=\sqrt{\frac{3}{N}\sum_{j=1}^{N} (-1)^{i+j}
\langle S^z_i S^z_j \rangle},
\label{staggmag.eq}
\eea
where the sum extends over all sites and the $SU(2)$ symmetry of the Hamiltonian
was used. The quantity $\mu^{QMC}(i)$ is the QMC estimate of the staggered magnetization at lattice sites $i$.
Measurements of this quantity for the octagonal tiling
are plotted
against the coordination number $z$ in Fig.\ref{staggmagQMC.fig} (data obtained
for the N=1393 approximant). The values shown were normalized with respect to the average value
 $\overline{\mu_s} = 0.337 \pm 0.002$ determined by extrapolation to
the thermodynamic limit.

\begin{figure}[h]
\includegraphics[scale=0.4]{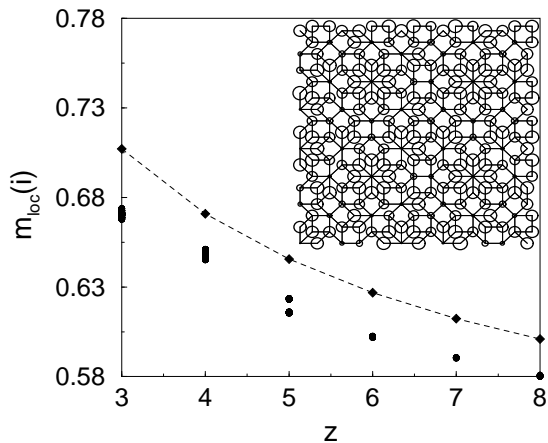}
\vspace{0.0cm}
\caption{
The local magnetization plotted against coordination number $z$.
The points on the dashed line are the exact values obtained for
Heisenberg stars.
The inset shows a portion of the tiling
where the radii of the circles correspond to the size of the local
magnetizations.
(Reproduced from \cite{wess})}
\label{staggmagQMC2.fig}
\end{figure}

As seen in Fig.\ref{staggmagQMC.fig}, local environments clearly play an important role.
To sort out the roles
of short-range  versus  long-range effects, one can introduce another estimator for the local staggered
magnetization,
\bea
\mu^{QMC}_{loc}(i)=\sqrt{\frac{3}{z+1}\left(\frac{1}{4}-\sum_{j=1}^{z} \langle S^z_i S^z_j \rangle\right)},
\eea
where the sum is truncated to the sites in the immediate
vicinity of $i$. This quantity is a local approximation of the one defined in
Eq.\ref{staggmag.eq}.
The results for this local estimator for the local magnetization is plotted in Fig.\ref{staggmagQMC2.fig} along with, for comparison, the prediction of the Heisenberg star calculation (Eq.\ref{heismag.eq}). The trend towards smaller values of staggered magnetization as $z$ is seen in both cases. One can conclude from these results that the nonmonotonic dependence of the full staggered magnetization $\mu^{QMC}_{loc}(i)$ as a function of $z$ is due to longer range correlations in the quasicrystal. This will be further explored in the section devoted to linear spin wave theory.

The global average of the staggered magnetizations for the octagonal tiling is
 $\mu_s = 0.337 \pm 0.002$  which is larger than the value found for the square lattice
($\mu_s = 0.3071 \pm 0.0003$\cite{sand}). This indicates that
quantum fluctuations reducing the order
parameter are suppressed in the quasiperiodic tiling
due to the inhomogeneous connectivity. The ground state energy per spin is $-0.6581$, compared to the value $-0.6694$ (square lattice) again showing that the square lattice has bigger quantum fluctuations than the quasicrystal.

For the Penrose tiling, QMC calculations \cite{sza} for the staggered moments show that, as for the octagonal tiling, staggered magnetizations are distributed in a range of values, with a strong coordination number dependence. These results are reviewed later on, when we compare the QMC and linear spin wave theory for the Penrose tiling antiferromagnet.

To explain the observed results one can turn to other theoretical tools such as renormalization group or linear spin wave theory, described next.

\subsection{Scale transformations and Renormalization Group}
We have mentioned that an important symmetry of the tiling, namely
invariance under discrete scale transformations (inflations/deflations) suggests the possibility of calculating properties by a renormalization group (RG). This method is used for problems with scale invariance, and has been successfully applied to systems near a critical point, for example.

 The renormalization group scheme described here is a generalization of the
calculation of Sierra and Martin-Delgado for the square lattice antiferromagnet
\cite{sierra}. In this approach \cite{jag,jagrg}, it was shown that under a change of scale, the original Hamiltonian is transformed into a new Hamiltonian of the same form but with rescaled parameters. The first step consists of defining blocks of 5 spins (formed by a central spin $S$ and its four nearest neighbors), such that the blocks form a square lattice of length that is $\sqrt{5}$ times bigger. The original spins of each block were next combined to form an effective block spin.
The important point is that the interactions between the block spins can be written in a form of a Heisenberg model with a new nearest neighbor coupling which depends on the old coupling, as well as the spin and length rescaling factors. The mapping is not exact, and among other approximations, some terms of the Hamiltonian are neglected. This calculation
for the square lattice was adapted to the octagonal tiling
in \cite{jagrg}. The case of the Penrose tiling, more complex, has not so far been attempted. For the octagonal tiling, the global ground state energy was computed, and in addition, values for local site-dependent
order parameters were obtained. The approach is briefly described below.

We have seen that the sites of the octagonal tiling are classified into A,B,C,D,E and F in order of decreasing coordination number (from $z=8$ for A sites down to $z=3$ for F sites). The frequencies (or probability of occurrence in the tiling) of each type of site is exactly known, in terms of the number $\lambda$.
Consider a finite piece of the tiling of $N$ sites. Under deflation, where the length scale is increased by $\lambda$, the  number of sites is reduced by a factor $1/\lambda^2$.  One can show that the sites which disappear are those having $z=3,4$ and half of the sites with $z=5$, called $D_2$ sites. The sites that remain are the A,B,C, and half of the D sites (the $D_1$ sites). They are transformed into the A,B,C,...,F sites of the new tiling and have a new value of the coordination number $z' \leq z$. For example, a B site always transforms into a $D_2$ site under deflation, a C site site into an E site, and so on as in Table 2. These four families of sites are used to introduce block spins variables.

\begin{center}
\begin{tabular}{|l c r|}
\hline
 initial site && final site \\
 (z value ) && (z' value) \\
\hline
A (8) &\quad \quad $\rightarrow$ \quad \quad &A,B,C or $D_1$ (8,7,6,5) \\
B (7) &\quad \quad $\rightarrow$ \quad \quad & $D_2$ (5) \\
C (6) &\quad \quad $\rightarrow$ \quad \quad & E (4) \\
$D_1$ (5)&\quad \quad $\rightarrow$ \quad \quad & F (3)\\
\hline
\end{tabular}
\vskip 0.5cm
\small{Table 1. Transformations of sites in the octagonal tiling under an inflation}
\end{center}

\begin{figure}[ht]
\begin{center}
\includegraphics[scale=0.60]{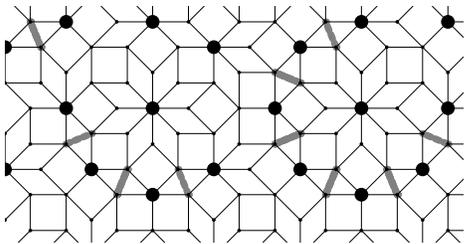}
\vspace{.2cm} \caption{Tiling showing block centers (black dots).
The grey lines connect pairs of sites that are shared between two
blocks. (Reproduced from \cite{jagrg}) } \label{decoup.fig}
\end{center}
\end{figure}

Each of the ``new" sites after deflation is associated with a given cluster of sites, comprising a central site and its $z$ neighbors, on the original tiling. There is however a first complication that was not present in the square lattice, due to the fact that, in the octagonal tiling, some of the clusters turn out to share edges. The first approximation made in the RG scheme is therefore to remove a small fraction, $\sqrt{2}/\lambda^3 \approx 10 \%$ of the bonds. This bond dilution allows decoupling of clusters. Fig.\ref{decoup.fig} shows the sites of the block spins by black dots, and the bonds that have to be taken out in order to obtain decoupled clusters. Note that the dilution of the OT leaves it two-dimensionality intact.

\begin{figure}[ht]
\begin{center}
\includegraphics[scale=0.8]{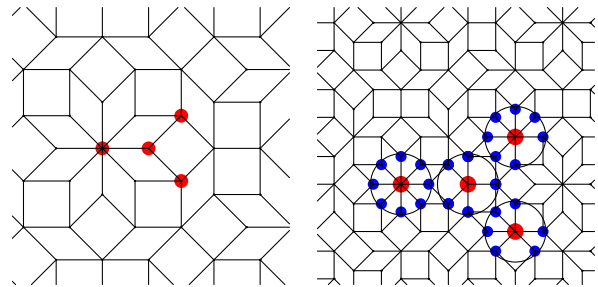}
\caption{(left) Example of an F-site($z$=3)(left) with the central and peripheral sites indicated by red dots. (right) Example of an F-site on a larger scale with block spins defined on clusters of sites as shown.(Reproduced from \cite{jagrg})}
\label{fclus.fig}
\end{center}
\end{figure}

One can now define a new ``block spin" variable $S' = (z-1)S$ composed of the old spins $S$ of the $z+1$ spins belonging in each of the clusters.  Fig.\ref{fclus.fig} shows an F-site (of coordination number $z=3$) on the original tiling, and, on a larger length scale, a composite F-site made up of a central block spin surrounded by three other block spins on the tiling. Upon repeating the inflation transformation, clusters can be defined on larger and larger length
scales. Notice that each of the four block spins of the figure on the right has a different internal structure -- this illustrates one of the differences that make the quasicrystal such a difficult problem compared to a periodic system. The next important approximation made in the RG is to restrict the number of such combinations possible, so that one works with a finite set of variables rather than the infinite number that is needed in principle for the quasicrystal. In the Heisenberg model, this truncation to a finite set of environments works because of the fact that next-neighbor and further-neighbor corrections decrease very quickly on the tiling. Contributions to the effective Hamiltonian of a given central spin due to far-off spins are quantitatively unimportant, and they will simply lead to a broadening of the block spin values obtained for each of the five types of blocks.

As in the square lattice, it can be shown that the block spins of the inflated tiling interact via an effective antiferromagnetic interaction. Just as in the case of the square lattice \cite{sierra}, one can write an approximate effective Hamiltonian
as a sum of
single-block contributions (the diagonal terms) and a set of terms
involving nearest neighbor blocks (off-diagonal terms). That is to say, the expression for the energy of the inflated system can be written so as resemble the original nearest neighbor Heisenberg Hamiltonian, up to  multiplicative factors and an additional term.
The renormalization group (RG) transformation relates the new and the old Heisenberg Hamiltonians, by specifying a) the set of transformations of spins $S$ into block spins $S'_j$ (where $j$ labels the type of block) and  b) the set of transformations of the old couplings $\{J\}$ into new effective couplings $\{J'_{ij}\}$. The number of spins is reduced from $N$ to $N'$. The formal expression for the transformed problem reads

\begin{eqnarray}
T^\dagger_0 H(N,S,J) T_0 \approx N'e_0(J,S) + H'(N',S',J') \nonumber \\
\end{eqnarray}

where, after the scale transformation $T_0$, the new Hamiltonian $H'$ has the same form (bilinear in
$S'$) as $H$.  The new problem that one encounters in a quasicrystal, as we have seen already, is that the number of types of blocks and the number of types of couplings between them increases with the number of steps of inflation. The approximate scheme described in \cite{jagrg} truncates the number of cluster types and cluster-couplings that are retained in the calculation to just seven cluster types and five inter-cluster couplings. The new couplings $J'_{ij}$ couplings are related by a multiplicative (matrix) relation to the old and and decays with increasing number of deflations. The block spins are similarly related by a matrix equation to the original spins and they increase with $n$. The resulting set of matrix equations for the couplings and the spins is then diagonalized to obtain the fixed point solution of the system. The solutions for the block spins flow to infinity, i.e. the classical limit, as in the periodic lattice, with a fixed ratio between the seven different block spins.

The ground state energy of a single block is that of a $z$-fold Heisenberg star (Sec.II.??) in which $z$
spins of lengths $S_j$  are coupled with
strength J to a central spin $S_0$. The total ground state energy $E_{0}$ can be written as the sum over all blocks of the block energies, at all
orders, as follows:

\bea E_{0}/N =  \sum_{i \in \alpha} f_i(\epsilon^{(0)}_i +
\frac{1}{\lambda^{2}}\epsilon^{(1)}_i + ... \frac{1}{\lambda^{2n}}
\epsilon^{(n)}_i + .....) \label{gse}\eea

The zeroth order contribution is the sum over cluster energies $\epsilon^{(0)}_i$ where $i$ labels the type of site on the original tiling, whose frequency of occurrence is $N^{(0)}(z)/N = f_i$. With each deflation, the number of
clusters decreases as $\lambda^2$. The energies after $n$ deflations, $\epsilon^{(n)}$, are the energies of blocks with a spin
$S_0^{(n)}$ at the center, with effective couplings
${\overline{\jmath}}^{(n)}$ to the $S_i^{(n)}$ surrounding spins.
Summing the series
for the energy gives $E_{0}/N \approx -0.51$, which lies about 20\% below the value deduced from numerical QMC data \cite{wess}. This order of magnitude of error is also found in the RG solution for the square lattice, and arises in part due to the neglect of certain types of bonds in this approximate scheme (see \cite{sierra}).

\begin{figure}[ht]
\begin{center}
\includegraphics[scale=0.7]{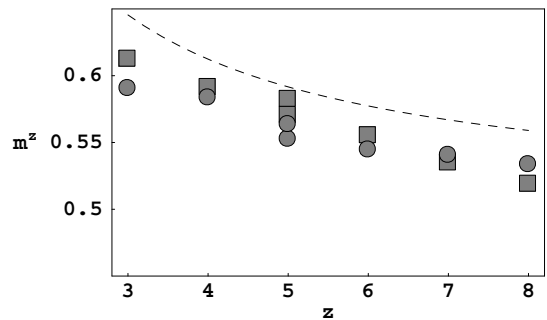}
\vspace{.2cm} \caption{Staggered magnetization (as a fraction of the full value $S$) obtained by the RG calculation $m_i$ at fourth order after correction for bond dilution (grey rectangles), the QMC data (grey circles)and the analytic zeroth order result (dashed line). (Reproduced from \cite{jagrg}) }
\label{staggmagsrg.fig}
\end{center}
\end{figure}

For the tiling, it is possible to calculate the relative strength of the local staggered magnetizations on each of the six families of sites by this RG method. These are determined by converting the cluster energies into estimates for the staggered magnetization using the same argument as in Eq.\ref{heismag.eq}. Fig.\ref{staggmagsrg.fig} shows the staggered magnetizations $m_i$ (as a fraction of the full value S)  for $z$ values ranging from 3 to 7 for a 4th order calculation after correcting for the bond dilution. Also shown are the zero order result (dashed line) and the QMC data.  The order parameter decreases with increasing $z$, agreeing qualitatively with Quantum Monte Carlo . The quantitative accord is also very good.

\subsection{The Linear Spin-Wave Approximation}
Spin wave theory gives the quantum corrections to classical ground state. It is only useful in situations when the exact ground state is known to be close, in a sense to be clarified below, to the classical state, in this case, the N\'eel two-sublattice up-down state. In the quasicrystal, as in other unfrustrated Heisenberg antiferromagnets, one expects there will be
long range order at zero temperature, as in the previously described sections.  In the ground state,
spins acquire a non-zero average value along the z-direction, with
 $m_{0si}=
\langle \epsilon_iS^{z}_i\rangle_0 \neq 0$ where $\epsilon_i =\pm 1$ depending on whether
the site is on the A or B sublattice. In spin wave theory, one assumes that the deviations from the classical value are small compared to the spin length, i.e. $S-m_{0si} <<S$, justifying a perturbation theory around the classical state.  In this case the exact Heisenberg Hamiltonian can be approximated by the first several terms of a power law expansion in spin deviation operators. Quantities such as magnetizations and bond strengths can then be calculated to successive orders of this expansion.

In the quasicrystal, the sites are expected to have different values of the local staggered magnetization. We assume that quantum
fluctuations are small for each of the sites taken separately, and carry out an expansion around the classical antiferromagnetic
ground state by introducing spin deviation operators for each site. According to standard definitions (see for example
\cite{manou}), this means introducing two families of Holstein-Primakoff boson operators: ${a_i,a^\dag_i}$ for sites on the A
sublattice, ${b_i,b^\dag_i}$ for sites on the B sublattice. The
boson operators account for deviations from the classical
configuration, with the spin deviation operator on A-sublattice
sites given by $a_i^\dag a_i = \widehat{n}_i \equiv
S-\widehat{S}_i^z$ and on B sublattice sites $b_j b_j^\dag =
\widehat{n}_j \equiv S+\widehat{S}_j^z $. Expanding the original Heisenberg Hamiltonian and keeping terms upto second order in the boson operators one obtains the linear spin wave (LSW)
Hamiltonian
\[H_{LSW} = -JS(S+1)N_b+JS\sum_{<ij>}(a_i^\dag a_i+b_j b_j^\dag+a_i^\dag b_j^\dag+b_j a_i)\]
where \(N_b = 2N \) is the number of bonds, with \(N\) being the number of
sites. This quadratic Hamiltonian can be diagonalized by a
generalized  Bogoliubov transformation \cite{white}. The
diagonalization is carried out in the real space basis, for finite systems -- contrarily to
periodic systems, where using a Fourier representation leads to significant simplification of calculations.

After the Bogoliubov transformation, which takes the set \{$a_i,b_j$ \}
to a set \{$\alpha_m,\beta_m$ \},  the
Hamiltonian takes the diagonal form :
\begin{equation}
  \label{eq:Hsw_diag}
H_{LSW} =  E_0 + \sum_{m=1}^{N/2} \Omega_m^{+} \alpha^\dag_m
\alpha_m + \Omega_m^{-} \beta^\dag_m \beta_m
\end{equation}
where $\Omega_m^{+}$ and $\Omega_m^{-}$ are the energies of the eigenmodes. In crystals the eigenmodes described by the $\beta$ operators are spin waves: they carry both a spin and a crystal momentum. In the quasicrystal, the eigenmodes are spatially much more complex. Depending on the energy of the mode, we will see that they can be more or less spatially extended, and for certain values of the energy, they can be strongly localized on a subset of sites of the tiling. Nevertheless, in the limit of very low energies, or very large length scales, the details of the quasiperiodic structure are not pertinent, and the solutions are expected to approach extended states. In this limit, the dispersion relation should resemble the usual linear relation found in antiferromagnets when $k\rightarrow 0$, $E(k) \approx ck$, along with a density of states $N(E) \sim E^2$ in $d=2$. This is indeed observed (see the section on the density of states).

$E_0$ is the ground state energy, comprising a classical part and a term arising from quantum corrections $\delta E$,
\begin{eqnarray}
E^{LSW}_0 &=& -JS(S+1)N_b +JS \sum_{m=1}^{N/2}\Omega_m^{-}  \nonumber \\
&=& -JS^2 N_b + \delta E
\label{gse.eq}
\end{eqnarray}
The local staggered magnetizations are calculated using the following expressions valid for sites $i$($j$) belonging to the A(B)-sublattice respectively:
\begin{eqnarray}
 \label{eq:lsm}
 m_s(i)=|\langle S^z_i \rangle |=S- \langle a^\dag_i a_i\rangle_0 \nonumber \\
 m_s(j)= |\langle S^z_j \rangle |=S-\langle b^\dag_j b_j\rangle_0
\end{eqnarray}
where the expectation values are calculated in the ground state.

\subsection{Finite size samples for numerical calculations.}
Calculations can be done for finite samples of two types. One can consider finite portions of an infinite system, with free boundary conditions. Such samples have
the usual problem of spurious boundary modes, which are difficult to correct for. Such samples can be generated by projecting down from five dimensions (in the Penrose tiling) or four dimensions (octagonal tiling) using the standard projection matrices. The second alternative is to consider approximants as in previous work on electronic (see review \cite{pmreview}) or phonon models \cite{los}. These, as we saw already are closely related to the infinite quasiperiodic structure at small length scales, but different on large length scales. For the Penrose tiling, one can generate rectangular shaped approximants obeying periodic boundary conditions called the Taylor approximants \cite{sza}. For the octagonal tiling, one can generate square periodic approximants \cite{ogu}. By considering the problem in periodic approximants one eliminates the problem of boundary effects, at the cost of introducing a global shear-type distortion. The latter has consequences for the energy spectrum as the approximant has a symmetry not present in the quasicrystal. It turns out that, fortunately, there is good agreement between the results for the two different kinds of system when one considers thermodynamic quantities  obtained by averaging over modes. For example, the order parameters for spins deep in the interior of finite samples have the same distribution as in the periodic approximants.

\section{Results for ground state energy and staggered magnetizations}
For the Penrose tiling, the ground state energy per site (GSE/site) is found from Eq.\ref{gse.eq} for
the periodic approximants of size N ranging from 96 to 11556.
The extrapolation to infinite
size gives an asymptotic value $E_0 = -0.643(0) \pm 0.0001$. For the octagonal tiling, the LSW calculation gives a ground state energy $E_0 = -0.646$ \cite{wess2}. The value for the square lattice is -0.658 (which decreases to -0.6705 when the next term in the perturbation series is taken into account, see \cite{manou}).

The ground state energy per site and bond is shown in Table 1 for different types of lattices. Of the three systems presented here, the lowest GSE/site is that of the square lattice, then the octagonal tiling, and, very slightly higher, the Penrose tiling. In addition, we show an approximant of the octagonal tiling (the ``crown" lattice), which has 7 sites per unit cell and was studied in \cite{jmw}. This approximant intercalates between the square lattcie and the octagonal tiling. The ground state of the Penrose tiling is closest to the classical ground state because, presumably, spin waves are ``harder to excite". These results should be treated with caution because they are obtained for approximant systems. The higher-order corrections in spin wave theory will play an important role in dtermining the final value of ther spin wave velocities. Nonetheless, based on the LSW results, the higher energy of the PT could reflect the slightly higher ``complexity" of the PT compared to the OT. One could expect, therefore, that the GSE/bond for tilings of high co-dimensions \cite{vidal2} will lie above that of the PT. The GSE/bond of the Rauzy tiling of Vidal and Mosseri \cite{vidmoss}, would, on the contrary, lie below that of the OT. These expectations have not been checked so far by calculations.

\begin{small}
\begin{center}
\begin{tabular}{|c|c|c|c|c|}
\hline
lattice & GSE  & GSE  &  bond  & GSE/bond   \\
 & (cl.) & (LSW) & (cl.) & (QMC) \\
\hline
Penrose &  -0.5 & -0.643 & -0.25 & -0.322 \\
Octagonal tiling & -0.5  &  -0.646 & -0.25& -0.323 \\
Crown lattice & -0.5  &  -0.648 & -0.324 & -0.330 \\
Square lattice& -0.5  & -0.658  & -0.25 & -0.335 \\ \hline
\end{tabular}
\vskip 0.5cm \small{Table 1. Classical and LSW values for ground state energy per site and per bond (for $S=1/2$ and in units of J) for the two quasiperiodic antiferromagnets, an approximant system (the crown lattice) and the square lattice}
\end{center}
\end{small}

\begin{figure}[t] \begin{center}
\includegraphics[width=8cm]{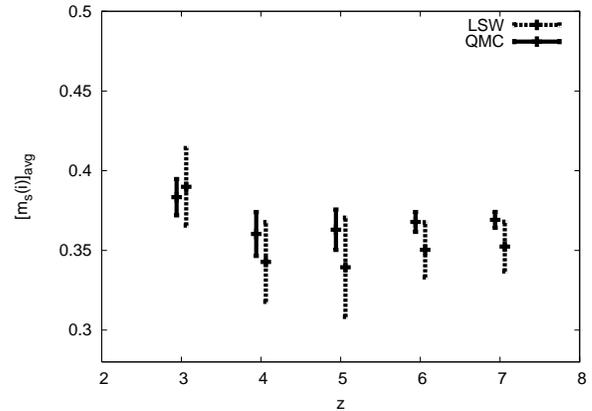}
\caption{\label{fig:mzfunc} Averages and standard deviations of the
local staggered magnetization as a function of $z$  ( LSW theory (dashed line) QMC (continuous line) for Taylor approximant N=4414).(Reproduced from \cite{sza3})
}
\end{center}\end{figure}

\begin{figure}[t] \begin{center}
\includegraphics[width=8cm]{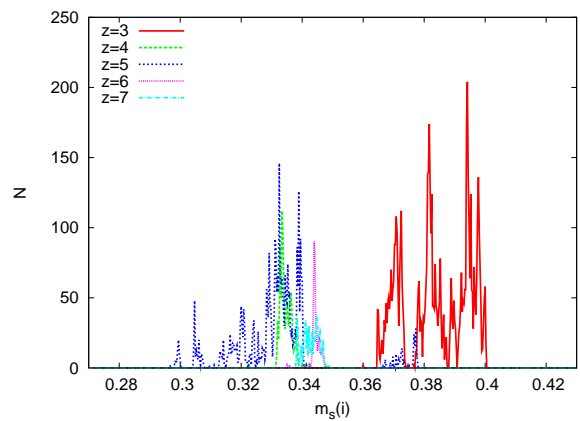}
\caption{\label{fig:distfn}  Probability distribution of the local staggered magnetizations for the five different values of $z$ in a Taylor approximant (N=11556).(Reproduced from \cite{sza3})
}
\end{center}\end{figure}

Fig.\ref{fig:mzfunc} indicates how values of the local magnetizations
vary with $z$ value, in the Penrose system (Taylor approximant size $N=4414$). The figure shows the
average values and the standard deviations of $m_i(z)$ for each
value of $z$, as obtained in LSW theory and from QMC ,
as reported in \cite{sza3}. The figure shows that Quantum Monte Carlo gives
a narrower spread of the magnetization. Linear spin wave theory clearly overestimates the fluctuations from average
behavior, giving a too high value for $z=3$ and too low values for higher values of $z$. The LSW theory tends to exaggerate the spread in values of the staggered magnetizations compared to QMC.
Fig.\ref{fig:distfn} shows the details of the $m_i(z)$
distribution for each of the $z$ values.
There is a continuum of values, as expected for a quasiperiodic
structure, with some pronounced peaks. Each of the main peaks correspond to specific next nearest configurations, as we will discuss next.

\begin{figure}[t] \begin{center}
\includegraphics[width=4cm]{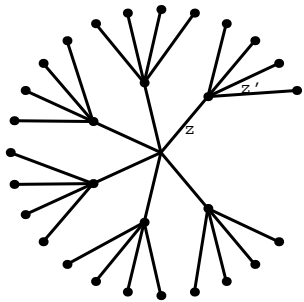}
\caption{\label{fig:hstar} Two-tier Heisenberg star showing $z$ near neighbors and $z'$ next nearest neighbors of the central site.(Reproduced from \cite{sza3})
}
\label{hstar2.fig}
\end{center}\end{figure}

In
order to explain the multiple peaks in the values of $m_{si}$ seen numerically in Fig.\ref{fig:distfn}
for a given $z$, we must take into account medium and long range structural details. This can be done analytically by considering clusters like the Heisenberg star cluster introduced already. As was shown in \cite{jmw}, it is easy to carry out a
LSW type analysis for the simple $z$-leg Heisenberg star in Fig.\ref{hstar.fig}, and calculate the magnetization of the central site as a function of $z$. To understand the results for the Penrose tiling, one needs to go a step further, and consider the effect of next nearest neighbors. This can be done by considering a two-tier Heisenberg star, as shown in Fig.\ref{hstar2.fig}. The LSW result obtained in \cite{sza2} showed that the staggered magnetization which now depends on $z$ and $z'$, $m_s(z,z')$ has a nonmonotonic behavior in $z$. The fact that the distribution of $m_s$ for the $z=3$ sites displays three peaks can be explained being due to the three different categories of these sites when one takes next nearest neighbors into account. Similarly, the $z=5$ sites fall into two categories due to the difference in $z'$, resulting in the splitting of the distributions into two groups for this class of sites.

\begin{figure}[ht]
\begin{center}
\includegraphics[width=5.0cm,angle=0]{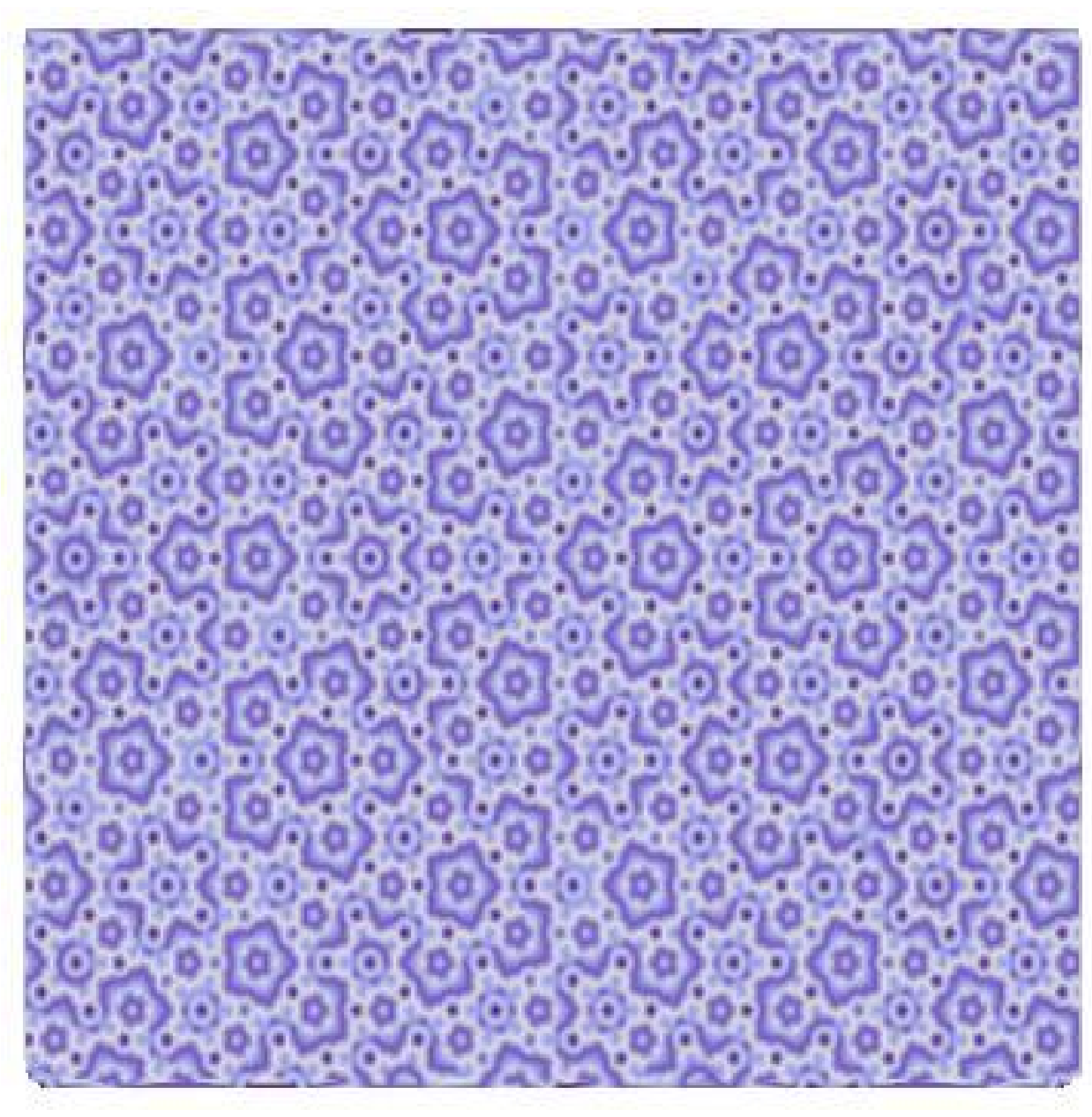}
\vspace{.1cm} \caption{Intensity plot of the staggered local magnetization $m_s$  in the Penrose antiferromagnet. Lighter and darker shades correspond to local values which are greater or smaller than the average value, respectively. }
\label{realsp.fig}
\end{center}
\end{figure}

Fig.\ref{realsp.fig} shows by an intensity plot how local magnetizations vary
in space on a portion of the Penrose tiling. The lowest values of staggered magnetization are found on a certain
subset of $z=5$ sites, while the largest values are found on the sites of $z=3$.

\subsection{Self symmetry of the staggered magnetization}
\begin{figure}[ht]
\begin{center}
\includegraphics[width=6.0cm,angle=0]{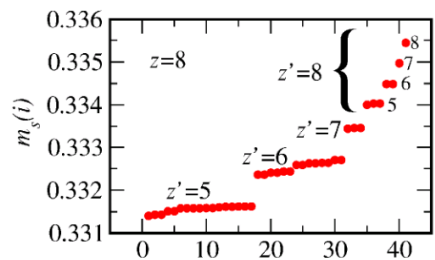}
\vspace{.1cm} \caption{Self-similar structure of the order parameter: the y-axis shows $m_{si}$ values of 40 A-sites of the octagonal tiling numbered (x-axis) according to increasing $m_{si}$ value. The values of the coordination number after one inflation ($z'$) and two inflations ($z''$) have been indicated. (Reproduced from \cite{wess2})}
\label{selfsim.fig}
\end{center}
\end{figure}

We have seen that the values of $m_{si}$ for a given $z$ are distributed about a mean value that depends on $z$, and that the actual shape of the distribution depends on details of the next nearest neighbors, and further nearest neighbors. In this section, we focus on the order parameters of a subset of sites which are self-similar under inflation and we show the resulting self-similar structure of the distribution of order parameters. These are the A-sites of the octagonal tiling, i.e. the sites with eight-fold symmetry and coordination number $z=8$. Calculations show that the $m_{si}$ values for A-sites are grouped into four disjoint groups, each of which corresponds to a different next nearest environment, as we already explained. The uppermost group of values corresponds to A-sites which transform under deflation into A-sites. The next lowest group of $m_{si}$ values correspond to A-sites which transform into B-sites ($z'=7$), etc. Now, as can be seen in Fig.\ref{selfsim.fig}, the uppermost group can be seen itself to be composed of four subgroups, of which the uppermost subgroup corresponds to A-sites which transform under deflation into A-sites. This substructure would continues infinitely in an infinite system but stops at the third level of the hierarchy in the small samples that were considered for the numerics.

\subsection{Perpendicular space representation}
\begin{figure}[ht]
\begin{center}
\includegraphics[width=5.0cm,angle=0]{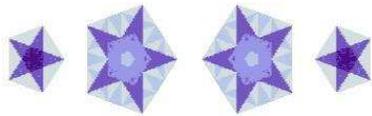}
\vspace{.1cm} \caption{ The same data as in Fig.\ref{realsp.fig} for the  ground state staggered magnetizations (ms) in perpendicular space. Each of the vertices of the Penrose tiling project into one of  four planes (see Box 3). The intensity at the point represents the value of  ms on that vertex.}
\label{perpsp.fig}
\end{center}
\end{figure}

It is interesting to consider the staggered magnetization from the viewpoint of the perpendicular space of the tiling \cite{sza3}. Recall that, in perpendicular space, the projections of different families of sites fall into separate domains.  When the $m_{si}$ values are plotted as a function of $\vec{r}_\perp$ rather than $\vec{r}_\parallel$ ( as was done in Fig.\ref{realsp.fig}) one obtains the figures shown in Fig.\ref{perpsp.fig}. The four domains shown lie at different distances along a third axis (not shown). Instead of the complex variations see in the real space representation, one sees five-fold star shaped domains, as illustrated in Fig.\ref{perpsp.fig}. The figures show that, as expected, the domain corresponding to each coordination number has a distinct color. The colors are not absolutely uniform since no two sites are identical.

Self-similar patterns can be investigated using this type of representation. To illustrate how this occurs, consider the octagonal tiling, whose behavior under inflation was discussed in the section on RG. If one considers the A (eight-fold) sites, for example, their transformation under inflation depends on the position in perpendicular space. The closer $\vec{r}_\perp$ to the origin, the bigger the number of inflations under which $A\rightarrow A$. Another way of saying this is that for such sites (i.e. those close to the origin of the perpendicular space window) the environment remains eight-fold symmetric out to greater distances. The staggered magnetization which depends on the environment, thus, depends on the perpendicular space coordinate, and should vary in magnitude progressively as one goes outward from the origin. To measure accurately this self similarity, first described in \cite{wess}, it requires calculations on systems large enough to permit a large number of inflations and to overcome the effects of distortions due to the boundary conditions.

To resume, the intensity plots in perpendicular space can serve not only to represent in an accessible and compact form the complex real space distribution of the order parameters on the quasiperiodic tiling, but also to demonstrate the properties of the state under length rescaling.

\subsection{Static structure factor}
The static longitudinal structure factor is defined by
\bea
{\mathcal{S}}^\parallel(\vec{k}) = \frac{1}{N} \sum_{i,j=1}^N e^{-i\vec{k}.(\vec{r_i}-\vec{r_j})} \langle S_{zi}S_{zj}\rangle
\label{static.eq}
\eea

\begin{figure}[t] \begin{center}
\includegraphics[width=5cm]{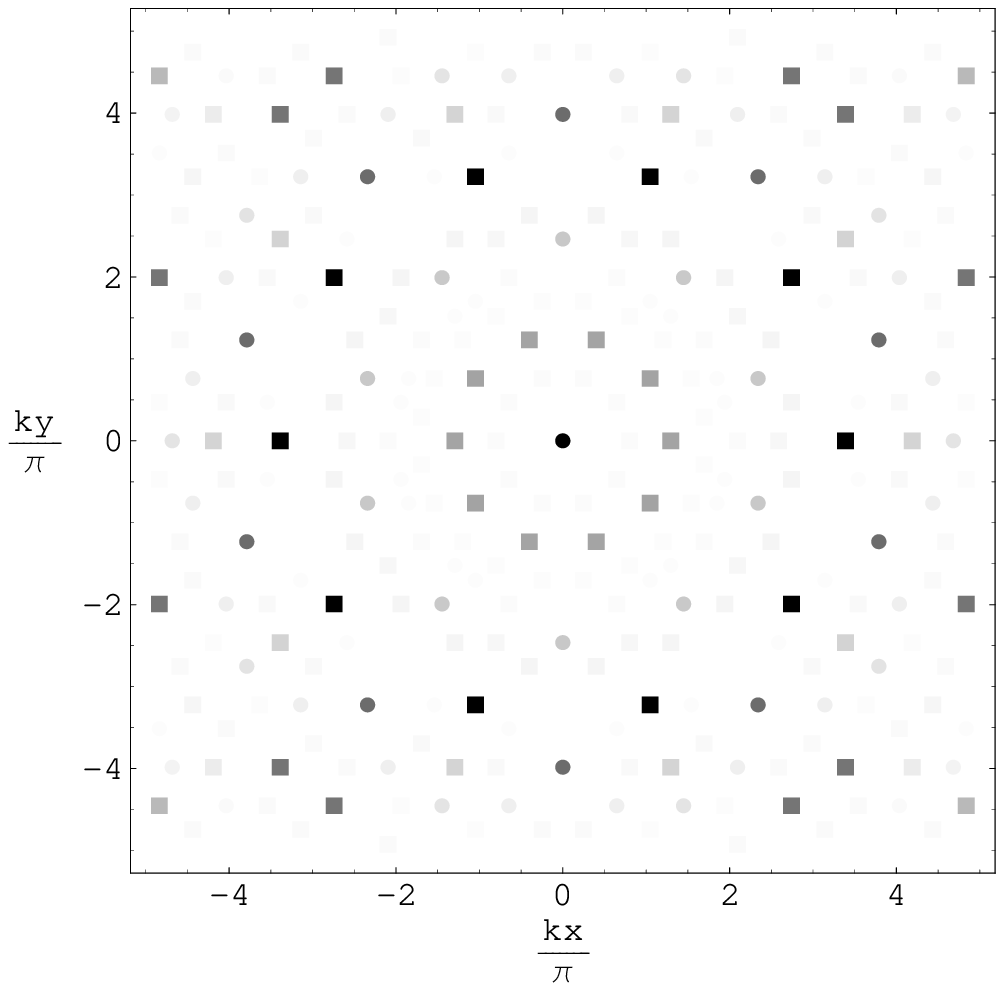}
\caption{\label{fig:fourier}
Intensity plot of the static longitudinal magnetic structure factor $S^{\parallel}({\mathbf k})$ for the $N= 4414$ Taylor approximant. The highest intensity peaks are shown in the figure at positions that are indicated by circles(squares) for the magnetic (nonmagnetic) structures respectively. The relative intensity is denoted by a linear gray scale ranging between zero (white) and maximum intensity (black). (Reproduced from \cite{sza3})
}
\end{center}\end{figure}

Fig.\ref{fig:fourier} shows an intensity plot of this static structure factor calculated for an approximant of the Penrose structure, for the nonmagnetic state and the antiferromagnetic ground state. There are many peaks of very low intensity of ${\mathcal{S}}$. The figure shows only the most intense peaks, whose positions are indicated by circles(squares) for the magnetic (nonmagnetic) structures respectively. The five-fold symmetry present in the magnetic state, as it is in the nonmagnetic state, along with the symmetry under $\vec{k}$ $\rightarrow -\vec{k}$ lead to the observed 10-fold symmetry of the peaks. The octagonal tiling antiferromagnet has a structure factor with 8-fold symmetry of both the magnetic and the nonmagnetic states. In both tilings, it can be seen that the magnetic peaks lie between the nonmagnetic peaks with no coincidences between the two. Peaks can be indexed using sets of four (or five) indices. These indices have integer values for non-magnetic peaks and can be integer or half-integer for the magnetic peaks. As in the case of the octagonal tiling \cite{wess2} the half integer indices arise  due to the doubling in size of the antiferromagnetic unit cell in five dimensions. This, coupled with the extinctions in the structure factor rule due to the multiplicity of each unit cell, then leads \cite{ron} to the observed ``shifting" or ``displacement" of the magnetic peaks with respect to the nonmagnetic ones. This shifting of indices would be true also for a three dimensional bipartite system, such as the icosahedral tiling, with spins interacting along edges of the rhombohedra. If all of the sites were to be occupied by spins, the antiferromagnetic state would give rise to half-integer indices as well in the structure factor. However, real life compounds such as ZnMgHo are more complicated.
This compound was studied by neutron scattering \cite{sato}, and seen to have short
 range antiferromagnetic correlations below about 20K. These
 correlations lead to a magnetic superstructure that is, as for
 our two dimensional model, shifted with respect to the
 paramagnetic state. The antiferromagnetic vector that best fits
 the data has a more complicated value than the simplest form for
 a 3d quasiperiodic antiferromagnet ($q_i = \frac{1}{2}, i =
 1,6$). This may occur because the magnetic unit cell is much larger for
 the three-component system, due to the fact that only the Ho
 sites carry a magnetic moment, and this results in smaller -- fractional -- distances between
 peaks in reciprocal space.

\section{Magnon spectrum and wavefunctions}
\subsection{Energy spectrum and density of states. Linear dispersion law.}
\begin{figure}[t] \begin{center}
\includegraphics[width=6cm]{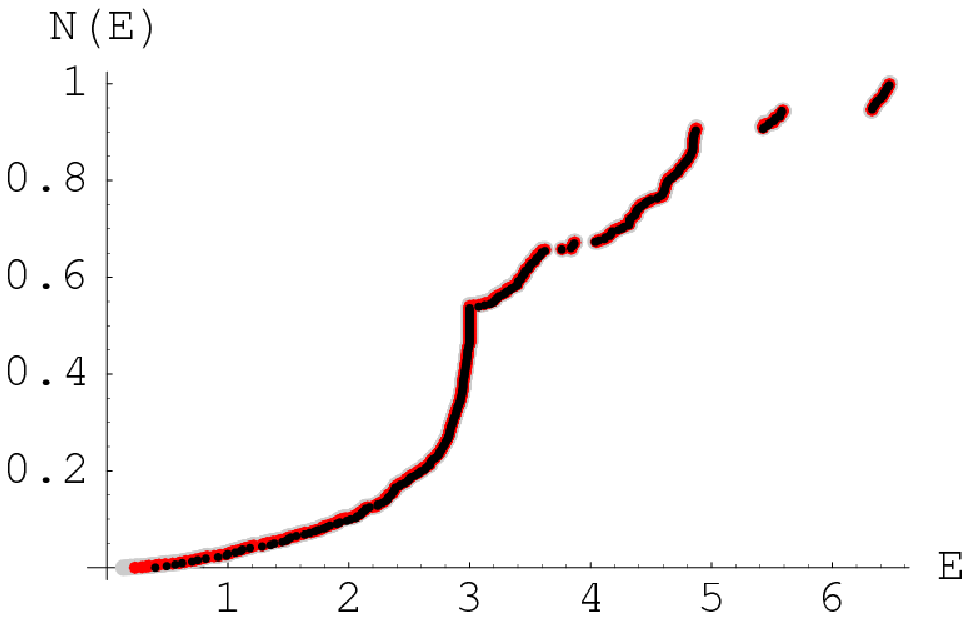}
\includegraphics[width=6cm]{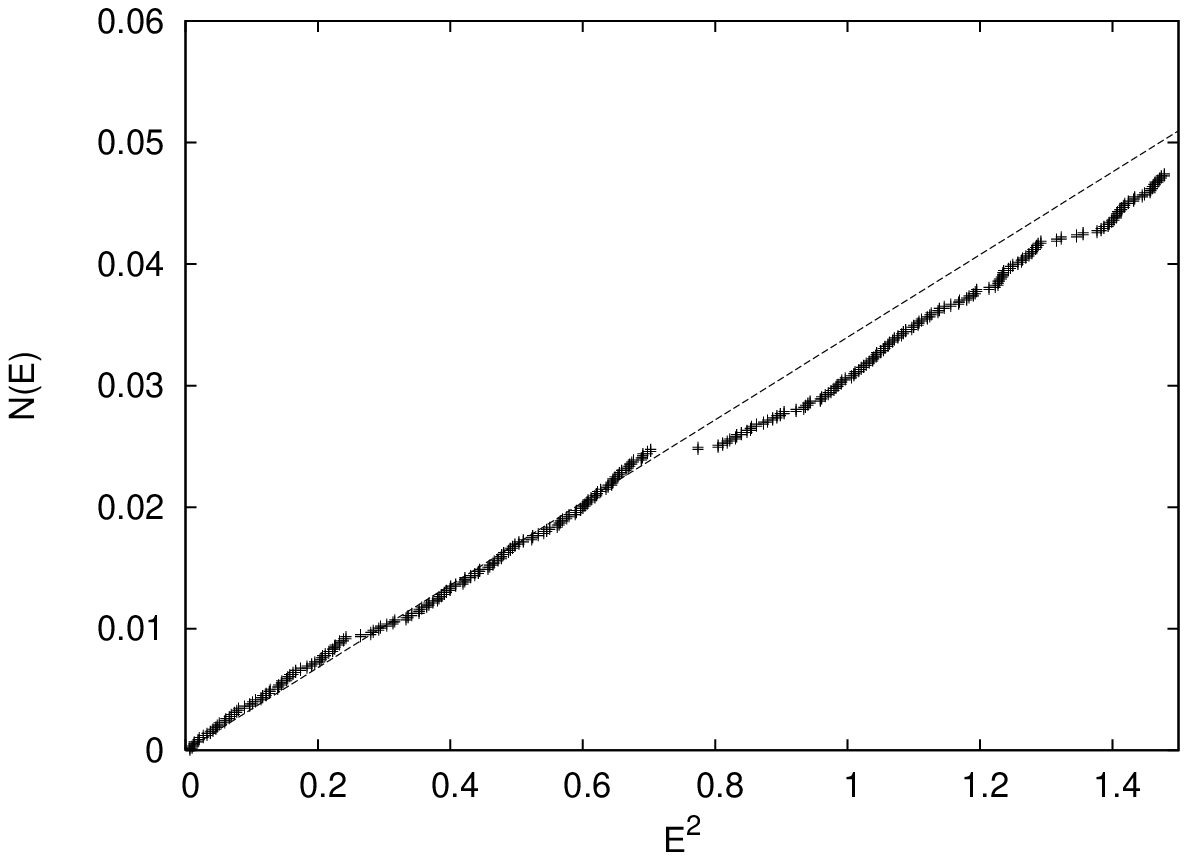}
\caption{\label{fig:IDOS_penrose} a)(Color online) Integrated density of states $N(E)$ versus $E$ (expressed in units of $J$) calculated
for three consecutive approximants ( black dots : $N=644$, red dots : $N=1686$, grey : $N=4414$) . b) Low energy tail of $N(E)$ versus $E^2$ (the straight line is a fit to the data).(Reproduced from \cite{sza3}) }
\end{center}\end{figure}

The low energy spin excitations, or magnons, have energies
$\Omega_m^{\pm}$. The two sets of energies $\Omega_m^{+}$ and $\Omega_m^-$
become identical in the limit of infinite size when sublattices A
and B become strictly equivalent.  The integrated density of states (IDOS) for the two sets of modes is defined by
\begin{equation}
N_\sigma(E) = \frac{1}{N}\sum_{m=1}^{N} \theta (E-\Omega_m^{\sigma})
\end{equation}
where $\sigma=\pm$.  The IDOS is plotted in Fig.\ref{fig:IDOS_penrose} for three different system sizes. The salient features of the figure are the
several groups of closely spaced energy levels, the main
gaps, which are stable with increasing system size, and a plateau at
the energy $E=3$, corresponding to a set of exactly degenerate states which will be discussed in the section on wavefunctions.

\begin{figure}[t] \begin{center}
\includegraphics[width=7cm]{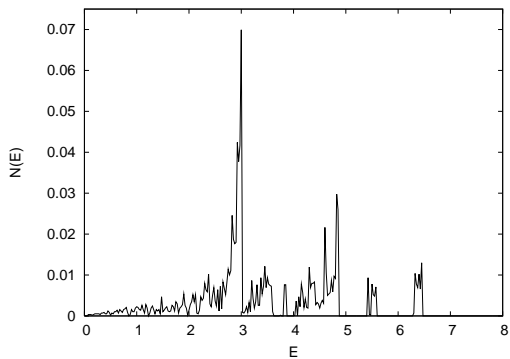}
\caption{\label{fig:DOS_penrose} Density of states calculated
for the Taylor 6 approximant (N=11556 sites).(Reproduced from \cite{sza3}) }
\label{dos.fig}
\end{center}\end{figure}

The derivative of the IDOS is shown in Fig.\ref{dos.fig}. One can distinguish several groups of energies separated by gaps. The local density of states (LDOS) allows one to determine which sites are most important at any given energy $E$. A study of the LDOS for the Penrose tiling system was carried out in \cite{sza3} and leads to the following conclusions :
1)The two highest energy bands around the values $E\approx 5.4$ and
$E\approx 6.4$, correspond to wavefunctions that are localized on
the 6- and 7-fold sites. 2) Lying below these
in energy are the states centered primarily on the 5-fold sites. 3) The
peak at the exact value of  $E=3$ corresponds to string-like states
living on the $z=3$ sites. 4) The lowest energies correspond to  the most extended states of the spectrum, having a two dimensional character and having its support on all types of sites.

As shown in Fig.\ref{fig:IDOS_penrose}b) the low energy part of the IDOS
can be fitted to a power law of the energy, $N(E)=\left(\frac{1}{8\pi c^2}\right) E^2$. This is consistent with
a linear dispersion of the magnon modes in this region of
the spectrum. Fitting to a form $N(E) = E^2/(8 \pi c^2)$ gives a
spin wave velocity on the Penrose tiling of $c = 1.1 J$. It is interesting to compare this result for $c$ with the corresponding values
on the square lattice and the octagonal tiling. On the octagonal tiling, the value is $c_{octa} \approx 1.3$J (Nb. a lower value of 1.1 J was obtained from a fit to the dynamical structure factor \cite{wess2}). The error in the spin wave velocities of the quasiperiodic systems is large, of the order of $10\%$, as the value depends on the region of the fit, and the system sizes that were considered are not big enough to allow a good smoothing of the low energy part of the IDOS. On the square lattice, the value can be calculated analytically in LSW theory, and it is
$c_{sq} =  2\sqrt{2}JS a \approx
1.41 J$ (for edge length $a=1$ and $S=1/2$).

To resume, excitations in the quasiperiodic antiferromagnets obey
an acoustic-type dispersion relation at long
wavelengths, as expected on general grounds for antiferromagnets. In LSW theory, the values of the spin wave velocity in the Penrose tiling is smaller than in the octagonal
tiling, which is in turn smaller than the value on the square lattice. For all systems, the edge lengths are all taken to be equal to unity. The propagation of modes appears to be slowest in the Penrose tiling. For the same edge length, one can calculate that the density of vertices $\rho$ (number of sites per unit area) is greatest for the Penrose tiling ($\rho$=1.2311), followed by that of the octagonal tiling ($\rho$=1.2071), followed by the square lattice ($\rho$=1). Since at long wavelengths the spin velocity depends on the mean density,
this provides an explanation of the trend observed for the sound velocities in these three systems.

\subsection{Wavefunctions.}
We now discuss the magnon wavefunctions and their spatial characteristics for
the different parts of the energy spectrum.

\begin{figure}[t] \begin{center}
\includegraphics[width=8cm]{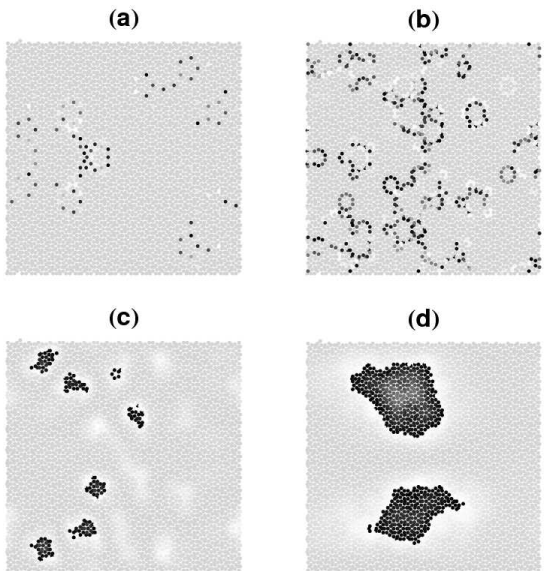}
\caption{\label{fig:wfnspar} Intensity plots in a PT approximant (N=4414) showing the
probability $\psi_i(E)^2$ for states of energy E (a darker shade corresponds to a higher probability): a) $E=6.469$ b)
$E=3.0000$ c) $E=0.8887$ d) $E=0.1692$. (Reproduced from \cite{sza3}) }
\end{center}\end{figure}

As seen by the large step in $dN(E)/dE$ at $E=3$ (Fig.\ref{dos.fig}), a large number of
degenerate states occur at this energy. This is due to wavefunctions
that have their entire support on the $z=3$ sites. These are string-like
states forming closed loops. The smallest loops are rings
around the ``football" $z=5$ sites. One typical linear combination of such degenerate states is shown in
Fig.\ref{fig:wfnspar}b).

The band of highest energies
corresponds to states that have large amplitudes on sites of $z=7$. The highest energy band has a width of about
0.16 and is centred around $E=6.4$.
The states involving $z=6$
sites correspond to energies in the range $5.43 < E < 5.49$.
These states have a smaller dispersion (flatter band) than the states in the
topmost group of energies. This may be explained by noting
that the six-fold sites are significantly fewer in number than the
seven-fold sites. These sites are therefore spaced further apart, which should lead to smaller wavefunction overlaps and smaller bandwidth.

The lowest energy states , for $E \leq 1$, correspond to wavefunctions for which the amplitude is non-zero on all categories of sites. The wavefunctions are in this sense more two dimensional than the previously mentioned wavefunctions. On the other hand, they are not truly extended -- they have large amplitudes within regions of size that depend on the energy $E$, and fall rapidly to zero outside these. It would be interesting to carry detailed studies of the behavior of these low energy states for large systems, in order to explore their multifractal properties. Figs.\ref{fig:wfnspar}c) and d) show the states
corresponding to two energies.

In disordered systems, a quantity that is often computed in order to determine the localized or extended character of wavefunctions is the inverse participation ratio (IPR), defined by

$$ P^{-1}(E) = \frac{\sum_j \vert \psi_j(E)\vert^4}{(\sum_j \vert \psi_j(E)\vert^2)^2} $$

The inverse of this quantity, or PR, corresponds, roughly speaking, to the number of sites on which the wavefunction is finite. For an extended state, this number is of the order of the system size N, while for a localized state the PR is a much smaller number -- only a few sites participate in the wave function. In general, the scaling of the IPR with system size is a gauge of the degree of localization (this is not true when there are macroscopic degeneracies, as in the string states of $E=3$). If the IPR varies as $P^{-1} \approx N^{-\beta}$,
then, as seen from the preceding remarks, $\beta=1$ corresponds to extended states, $\beta=0$ to localized states and intermediate values of beta to other cases including the so-called critical states.

\begin{figure}[t] \begin{center}
\includegraphics[width=8cm]{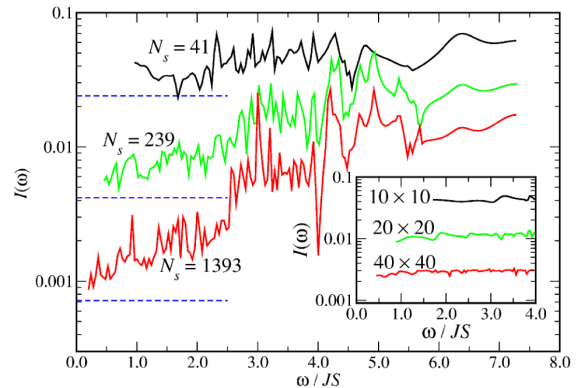}
\caption{\label{ipr.fig} The inverse participation ratio for states in OT approximants plotted as a function of energy ($\omega/JS$) for three different system sizes. Inset: the IPR as a function of energy for square lattice samples of different sizes, as indicated. (Figure reproduced from \cite{wess2})}
\end{center}\end{figure}

Milat and Wessel have calculated the IPR on three approximants of the octagonal tiling, and these are plotted as a function of the energy in Fig.\ref{ipr.fig}. One sees that, at low energies, the IPR depends on the system size, an indication that the corresponding states are extended. At high energies, the IPR flattens out and less size dependent, indicating that states are more localized. The inset shows, for comparison, the IPR calculated for square lattice samples, in which one knows the eigenstates to be Bloch states and therefore perfectly extended.

The studies of the IPR on approximants of the Penrose tiling \cite{sza3} confirm the observation that states are more delocalized at low energy compared to higher energies. The value of $\beta$ are largest in the low energy end of the spectrum, where it has an average value of slightly less than 1 ($\beta = 0.9\pm 0.1)$. The multifractal properties of the eigenstates can be investigated by calculating the function $f(\alpha)$ which contains information on the scaling of the wavefunction -- $f(\alpha)$ shrinks to a single point for a simple fractal or completely extended state but has a typical convex shape for Cantor set like wavefunctions, such as those at the metal-insulator transition \cite{grussbach}. There is some evidence for a nontrivial $f(\alpha)$ spectrum \cite{sza3}, and thus multifractal states, in the low energy region of the spectrum.

The figures showing
wavefunctions in real space can all be converted to perpendicular space representations \cite{sza3}, where there is a change, from localized to delocalized, of states as a function of $E$. This provides an alternative way of probing the wavefunction properties.

To sum up, we have shown some representative examples of magnon wavefunctions for the Penrose tiling approximants. Magnon wavefunctions have not been studied in sufficient detail, in large enough samples, however. Another problem that we have already mentioned in the context of the energy spectra is that relating to the boundary conditions. The wavefunction calculations shown here were done on a Taylor approximant of 4414 sites, with periodic boundary conditions. It is clear that the boundary conditions will have significant effects on the wavefunctions, particularly at small energies, where the wavefunctions are the most extended, although all other eigenmodes are in principle affected as well. This important problem has not been addressed satisfactorily thus far in any of the numerical calculations on tilings, be it for spins or electronic models.

\subsection{Dynamical structure factor}
\begin{figure}[t] \begin{center}
\includegraphics[width=8cm]{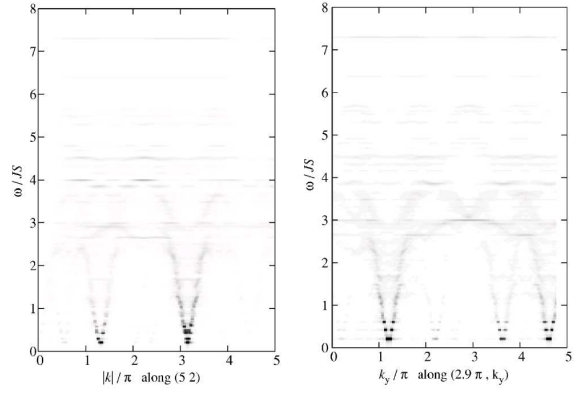}
\caption{\label{softmode.fig} Intensity plot of the dynamical structure factor of an OT approximant (N=1393) in the vicinity of magnetic Bragg peaks (left) along the line joining the peaks at $\vec{k}_B={1.2 \pi,0.5 \pi}$ and $\vec{k}_{B'}={3 \pi,3 \pi}$ and (right) along a line parallel to the $k_y$ axis and traversing $\vec{k}_B={2.9 \pi,4.8 \pi} $(right) (Figure reproduced from \cite{wess2})}
\end{center}\end{figure}

The dynamical magnetic structure factor is defined by
\bea
{\mathcal{S}}^\perp(\vec{k},\omega) =  \frac{1}{N}\sum_{i,j=1}^N e^{i(\omega t-\vec{k}.(\vec{r_i}-\vec{r_j}))}{\mathcal{S}}^\perp(i,j,\omega)  \nonumber \\
{\mathcal{S}}^\perp(i,j,\omega) = \int_{-\infty}^\infty dt \langle {\bf{S}}_{\perp i}(t).{\bf{S}}_{\perp j}(0)\rangle
\eea
where ${\bf{S}}_{\perp i}={S_{xi},S_{yi}}$ denotes the transverse components of the spin ${\bf{S}}_i$. For $\omega=0$, the dynamical structure factor reproduces the magnetic peak structure of the static structure factor already defined in Eq.\ref{static.eq}. The $\omega(k)$ can be measured by scans in $\vec{k}$-space. Figs.\ref{softmode.fig} show the intensity of the structure factor for scans effected around different magnetic Bragg peaks. The figure on the right shows the intensity profile in the $\vec{k},\omega$ plane as one varies $\vec{k}$ along the line joining two intense magnetic Bragg peaks at ${1.2 \pi,0.5 \pi}$ and ${3 \pi,3 \pi}$. One can see the soft modes, and linear dispersion near these two peaks. A fit to the data gives a spin wave velocity $c$ of about 2.1, in accord with the value that was extracted from the fit to the low energy end of the IDOS.

\subsection{Local frequency-dependent susceptibilities}
\begin{figure}[t] \begin{center}
\includegraphics[width=8cm]{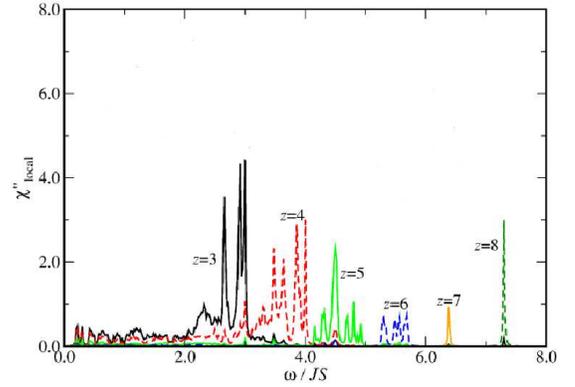}
\caption{\label{suscep.fig} Imaginary part of the local dynamic susceptibility plotted as a function of frequency $\omega$ averaged separately over sites belonging to six different $z$ values on the OT (Figure reproduced from \cite{wess2})}
\label{suscept.fig}
\end{center}\end{figure}

The dynamical structure factor yields information on the imaginary part of the local susceptibility via the relation
\bea
\chi"(i,\omega) = {\mathcal{S}}(i,i,\omega)
\eea
This quantity, which can in principle be probed by a nuclear magnetic resonance (NMR) experiment, was calculated for the OT approximant. Fig.\ref{suscept.fig} shows the six different curves obtained by averaging the local susceptibilities for all sites of a given value of $z$. As can be seen, all of the sites have a low frequency tail, but the different types of site have sharp peaks at energies that increase as $z$ increases. This provides a precise way of studying local environments: by tuning the frequency, one can probe different families of sites on the quasicrystal.

\section{Geometrically disordered quasiperiodic quantum antiferromagnets}
\begin{figure}[h]
\begin{center}
\includegraphics[scale=0.2]{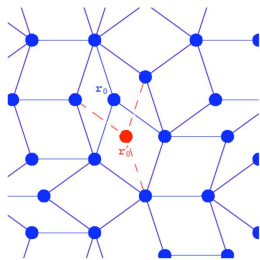}
\caption{(Color online)
A single phason flip showing the original ($r_0$) and final positions ($r'_0$, connected with rugged links). The original sites and bonds are shown in blue and the new site and new bonds in red.(Reproduced from \cite{sza4})}
\label{phasonflip.fig}
\end{center}
\end{figure}

In this section we consider the effects of geometric disorder due to localized phason flips. By ``phason flip" we will mean a local change of connectivities as shown in Fig.\ref{phasonflip.fig}, where a site in the interior of moves slightly from its original position. This has the effect of modifying the bonds locally but leaving the rest of the tiling unaffected. In addition, the bond lengths and angles remain unchanged, as the final result of the phason flip is simply a local rearrangement of the tiles. Many quasiperiodic structures display this type of phason disorder. The phasons can be quenched (static) in or they may be excited at finite temperature (dynamic). In the present context, we consider that the phason disorder is quenched into the system (the bonds do not fluctuate between configurations), and we are interested in the effect of this type of disorder on the antiferromagnetic ground state properties.

The consequences of this type of disorder are, as we will see, very different from the consequences of the more commonly considered models of disordered magnets, in which the couplings or the site occupations are assumed to be random. In the case of random couplings \cite{laflor}, the antiferromagnetic ground state is weakened progressively and ultimately destroyed as the amplitude of the fluctuations of bond strengths increases. In the case of random site dilution \cite{randomsite}, the antiferromagnetic state is weakened and a percolation transition can occur.
The calculations are carried out on samples with weak phason disorder. By this we mean that the starting point is a perfectly ordered approximant of the quasiperiodic tiling (the Penrose or the octagonal tiling) and one randomizes these by flipping a certain number of sites so as to create a set of disordered samples. The amount of phason disorder introduced in this manner is necessarily bounded, as opposed to the random tilings described, for example, in \cite{rantil}.

\begin{figure}[h]
\begin{center}
\includegraphics[scale=0.5]{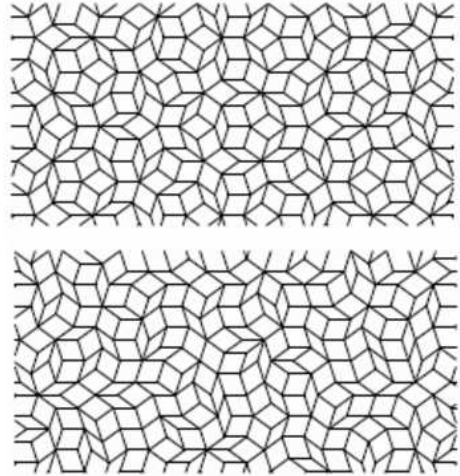}
\caption{Samples of (up) a perfect tiling and (down) a randomized tiling.(Reproduced from \cite{sza4})}
\label{tilings.fig}
\end{center}
\end{figure}

As Fig.\ref{phasonflip.fig} shows, the site undergoing a phason jump changes its sublattice. This is taken into account in the random selection of flipped sites, since the LSW and QMC calculations are performed on samples with an equal number of spins on each sublattice, $N_A=N_B$. Fig.~\ref{tilings.fig} shows a portion of a perfect(deterministic) tiling and a typical example of a disordered tiling after a certain numbers of phason flips. For a given total number of flips $N_{ph}$, the degree of disorder $\Delta$ is calculated by calculating the average overlap distance between the perfect sample and the disordered samples.

The model considered in \cite{sza4} is the unfrustrated Heisenberg model for spin-$\frac{1}{2}$, described by $H=J \sum_{\langle i,j \rangle} {\mathbf S}_i \cdot {\mathbf S}_j$ for spins situated on vertices of approximants of the Penrose tiling. Nearest-neighbor interactions are antiferromagnetic $J>0$, and act between pairs of sites that are linked by an edge. The model is disordered, with purely geometric disorder, i.e. the couplings $J$ acting along bonds remain fixed at a constant value. This is to be contrasted with a situation where couplings are random, for a given fixed connectivity of sites, as in the random exchange antiferromagnet on the square lattice \cite{laflor}.

\begin{figure}[h]
\begin{center}
\includegraphics[scale=0.6]{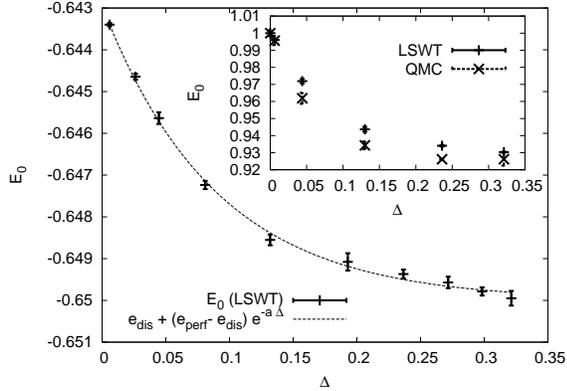}
\caption{The dependence of the ground state energy on disorder $\Delta$ calculated in LSWT. The smooth curve is a fit to an exponential form (see text). The system size is N=4414. The inset shows the normalized ground state energy as a function of increasing disorder $\Delta$. Points represent values obtained after finite size scaling of LSWT and QMC data. (Reproduced from \cite{sza4}) }
\label{GSE.fig}
\end{center}
\end{figure}

\begin{figure}[h]
\begin{center}
\includegraphics[scale=0.6]{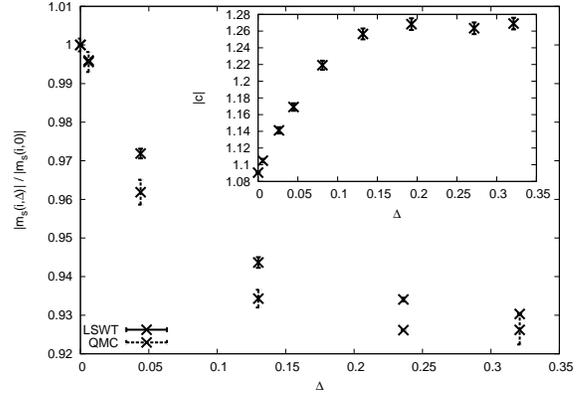}
\caption{Normalized average staggered magnetization dependence from the phason disorder $\Delta$. Points  represent values obtained after finite size scaling of LSWT and QMC data. The inset shows the effective spin wave velocity as a function of disorder $\Delta$ for several system sizes.(Reproduced from \cite{sza4})}
\label{avgmag.fig}
\end{center}
\end{figure}

\begin{figure}[h]
\begin{center}
\includegraphics[scale=0.8]{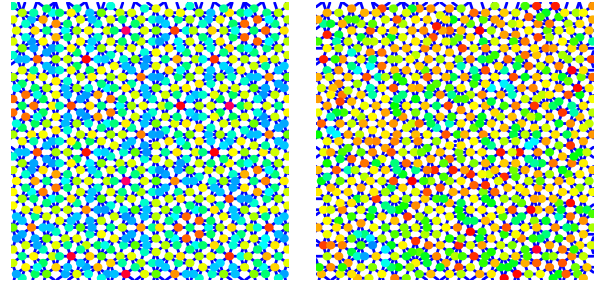}
\vspace{.2cm} \caption{(Color online) Color representation of the local staggered magnetizations in a perfect tiling (left) and a disordered tiling (right) showing sites of magnetizations ranging from high (blue) to low (red) values.(Reproduced from \cite{sza4})} \label{compare.fig}
\end{center}
\end{figure}

The results for the sample averaged ground state energy of the phason disordered Penrose antiferromagnet are shown in Fig.\ref{GSE.fig}. The values have been normalized with respect to the value obtained for the undisordered system. The results obtained from LSWT and QMC are in good agreement. This also shows the applicability of the linear spin wave approximations to this randomized system. The
smooth curve shows a fit to the form $e_0 = e_{dis} + (e_{per}- e_{dis}) e^{-a\Delta}$, with $a=11.16$, $e_{dis}=-0.6500$ and $e_{per}=-0.6429$. The asymptotic value of the GS energy $e_{dis}$ represents the average value of the ground state energy of maximally randomized Penrose approximants, which lies below the GS energy of the perfect system. In other words, overall, the nearest neighbor bond energy increases in absolute value. This indicates that the introduction of phasons tends to increase quantum fluctuations in the tilings. This is borne out in the calculations of the average staggered magnetization which should correspondingly decrease when disorder increases. This is indeed seen in Fig.\ref{avgmag.fig}, where the average staggered magnetization decreases as a function of the disorder strength. Fig.\ref{compare.fig} shows the magnetization profile by an intensity profile, with the perfect tiling shown on the left and a moderately disordered tiling on the right, for comparison. The colors in the figure on the right can be observed to be shifted towards the red end of the spectrum, indicating globally smaller values of $m_{si}$.

The low energy tail of the IDOS (integrated density of states) is quadratic, and can be fitted
to give a spin wave velocity that {\it increases} with $\Delta$ as shown in the inset of Fig.~\ref{avgmag.fig}. This indicates that spin wave propagation is facilitated by the
phason disorder, in analogy with the problem of quantum diffusion of
electrons in the tight binding model in quasiperiodic tilings
\cite{ben2}. In addition, one notices that the peak corresponding to localized states at $E=0$ disappear
progressively. These states arise, as shown in \cite{sza3} on closed
loops of 3-fold sites. They are destroyed when a phason flip
occurs on the one of the participating sites.

In conclusion,
geometrical disorder has important effects on the magnetic properties of a quasiperiodic
antiferromagnet. The real space properties of the ground state staggered magnetization profile are modified significantly. In particular, the hierarchical structure and five-fold symmetries are lost. The overall order parameter is smaller, due to the fact that there are more quantum fluctuations compared to the perfect tiling. It was also shown in \cite{sza3} the effective spin wave velocity for low energy modes is slightly higher, showing that disorder ``helps'' propagation of spin waves. Real space properties such as the spatial modulation of the order parameter are smoothed, and as regards the energy spectrum, the fluctuations of the density of states are smoothed. Degenerate states localized on closed loops disappear with increasing disorder. Eigenmodes tend to
be more spatially delocalized compared to the perfect tiling. All these effects are peculiar to quasicrystals, for example, the delocalization of eigenstates due to phason disorder has also been observed in phonon \cite{los} and tight binding electronic models \cite{pmreview}. These results, obtained for systems with an upper bound on the disorder, are not expected to necessarily hold when one allows for unbounded phason disorder. In the limit of strong disorder, antiferromagnets on random tilings in high codimension \cite{codimhi} are also interesting systems for future studies.

\section{Discussion and Conclusions}
The quantum antiferromagnet for quasiperiodic tilings without frustration has been described using different theoretical methods. Since the antiferromagnetic coupling was assumed to be the same for all the pairs of spins, the complexity of the Hamiltonian in the tilings arises solely from the aperiodic geometry. Two tilings were considered in particular: the Penrose rhombus tiling and the octagonal tiling, both with and without disorder. We compare the results for the antiferromagnets in these tilings with the square lattice antiferromagnet, which has, overall, the same number of bonds per spin and therefore the same classical ground state energy.

As in the square lattice, the ground states of the tilings have long range magnetic order, of the N\'eel type. Unlike the lattice, the local staggered magnetization $m_{si}$ is spatially inhomogeneous, and has a value that depends on the type of site $i$. In the perfect undisordered tilings, the values of $m_{si}$ depend primarily on the number of nearest neighbors but also on further near neighbor shells. The inflation symmetry of the octagonal tiling allows a calculation of the ground state properties by an approximate renormalization group. The results agree with the results obtained by Quantum Monte Carlo for this tiling. The main result of these calculations was to obtain the ground state energies of the quasiperiodic tilings and show that they are higher in energy for the periodic square lattice, which has the same number of spins and bonds. This is the first indication that the aperiodicity leads to a suppression of quantum fluctuations. The staggered magnetizations are in keeping with this conclusion, as the average value of $m_s$ is higher in the quasiperiodic tilings than on the square lattice.

To investigate these antiferromagnets in more detail, a linear spin wave theory was presented, and solutions obtained by numerical diagonalization. The results for the staggered magnetization distribution agree with that found by other methods. The dependence of $m_s$ on the local environments was investigated in some detail, and explained using a Heisenberg star cluster argument. The staggered magnetization in real space was seen to be a complex but highly symmetric function which is simplified when represented in perpendicular space coordinates. The Fourier transform of the magnetic state can be described by a quasiperiodic peak indexing and is related to the original nuclear structure factor by a generalized (i.e. in dimensions greater than the physical $d=2$) wave vector shift. This feature, characteristic of standard two sublattice antiferromagnets, has its counterpart in real three dimensional magnetic quasicrystals, as measured in neutron scattering experiments.

The spin excitations on the tilings have complex spatial structure, and show multifractal scaling. Their spatial extension is highly dependent on the energy, with the states of low energy being the most extended as seen by the scaling of the IPR (Inverse Participation Ratio) with system size. The correlations are long ranged.  Spin wave velocities, as found from fitting to the low energy end of the magnon spectrum decrease with increasing complexity of the tilings. The local dynamical susceptibility is shown to have sharp peaks as a function of frequency, and depends sensitively on the local environment. Thus, magnetic response could be a probe for self-similarity of quasiperiodic structures. The entanglement properties of such systems should present interesting differences with respect to the periodic lattice \cite{entangl}. Interesting nontrivial behavior of entanglement entropy has been found in one dimensional aperiodic spin models \cite{igloi} but two dimensional quasiperiodic spin models have not been studied so far.

Finally, the effects of phason flip disorder were described. The ground state increases as disorder increases, in the weak disorder limit. This leads to the conclusion that phason disorder makes the system more classical -- a conclusion also borne out by the fact that magnetizations globally increase in value. This is the opposite to the behavior usually found in crystals, where disorder tends to diminish the antiferromagnetic order parameter. Wavefunctions tend to become more extended, when phason disorder is increased -- this behavior was also observed in the tight-binding model for electrons in quasicrystals.

The two tilings studied are the best known of the two dimensional quasicrystals. Other tilings that could be investigated, in order to better understand the quantum fluctuations in aperiodic systems and the effects of disorder, are the 2d Rauzy tilings, and the rhombus tilings of Vidal et al. We have discussed unfrustrated models, on bipartite tilings. However in real quasiperiodic compounds, it is more likely that the structures are not bipartite, leading to frustrated antiferromagnetic interactions. This type of problem was recently discussed for classical spins \cite{motz} but has not been investigated for quantum models.

In conclusion, quasiperiodic antiferromagnets are a example of highly complex order in a magnetic system. Experimentally, the Heisenberg model may be possible to realize in thin films grown by deposition on quasicrystal surfaces. Cold atoms provide an alternative laboratory to test the theoretical predictions.

\subsection{Acknowledgments}
I thank Michel Duneau for many useful discussions over the years on geometrical aspects of quasiperiodic tilings. I also thank Stefan Wessel, Attila Szallas and Roderich Moessner for fruitful discussions and collaborations.

\end{document}